\documentclass{article}
\usepackage[labelfont=bf,labelsep=period]{caption}
\usepackage{graphicx}
\usepackage{subcaption}
\usepackage{PRIMEarxiv}
\usepackage{float}
\usepackage{placeins}
\usepackage[utf8]{inputenc} 
\usepackage[T1]{fontenc}    
\usepackage{hyperref}       
\usepackage{url}            
\usepackage{booktabs}       
\usepackage{xcolor}         
\usepackage{newunicodechar}
\newunicodechar{，}{,}
\usepackage{multirow}
\usepackage{threeparttable}
\usepackage{amsmath}        
\usepackage{tabularx}
\usepackage{amsfonts}       
\usepackage{nicefrac}       
\usepackage{microtype}      
\usepackage{lipsum}
\usepackage{fancyhdr}       
\usepackage{graphicx}       
\usepackage{titlesec}
\usepackage{indentfirst}
\usepackage{hyperref}
\usepackage{cleveref}

\titlespacing*{\section}{0pt}{1.5ex plus 1ex minus .2ex}{1ex}
\titlespacing*{\subsection}{0pt}{1ex plus .5ex minus .2ex}{0.5ex}

\graphicspath{{media/}}     

\title{CSU-PCAST: A dual-branch transformer framework for medium-range ensemble precipitation forecasting 
}

\author{Tianyi Xiong$^1$, Haonan Chen$^{1,2}$, Kelly Mahoney$^2$, Jingyin Tang$^1$, Tim Smith$^2$, Janice Bytheway$^2$\\
$^1$Colorado State University, Fort Collins, CO\\
$^2$NOAA Physical Sciences Laboratory, Boulder, CO\\
\thanks{\textit{Corresponding author: Haonan Chen (haonan.chen@colostate.edu)}}
} 

\begin{document}
\maketitle
\begin{abstract}
Accurate medium-range precipitation forecasting is important for hydrometeorological risk management and disaster mitigation, yet remains challenging for current numerical weather prediction (NWP) systems and data-driven models. Traditional ensemble forecasting systems, such as the Global Ensemble Forecast System (GEFS), can struggle with precipitation intensity, spatial structure, and ensemble reliability, especially at longer lead times. To address these challenges, this study develops CSU-PCAST, a deep learning-based ensemble forecasting framework for global precipitation prediction. The model is trained using fifth-generation ECMWF (ERA5) atmospheric and surface variables at \(0.25^\circ\) spatial resolution, with precipitation labels derived from NASA's Integrated Multi-satellite Retrievals for Global Precipitation Measurement (IMERG). CSU-PCAST uses 57 prognostic atmospheric and surface variables, together with static geographical fields, and predicts both the evolving atmospheric state and 6-h total precipitation. The proposed architecture employs a patch-based Swin Transformer backbone with periodically padded residual convolutions for longitudinal continuity, stochastic noise conditioning, time embeddings, and a dual-branch decoder that separately predicts precipitation and non-precipitation variables. During inference, CSU-PCAST is initialized from operational Global Forecast System (GFS) analyses and generates 30 ensemble members out to 15 days using an autoregressive strategy, matching the operational GEFS ensemble size. Evaluation against GEFS over the full year of 2023, using IMERG as the precipitation reference, shows that CSU-PCAST improves deterministic precipitation skill at short lead times, with higher CSI and lower RMSE during the first several forecast days. CSU-PCAST also reduces the strong GEFS wet bias for light precipitation and the dry bias at the 10 and 20 mm thresholds. Probabilistic verification shows lower CRPS, higher Brier Skill Scores for several thresholds, and improved ensemble reliability relative to GEFS, although both systems remain underdispersive. A case study of the Sanba extreme precipitation event further shows that CSU-PCAST provides improved spatial structure and exceedance-probability guidance, while also highlighting remaining challenges in regional precipitation totals and high-end precipitation intensity. These results indicate that CSU-PCAST provides improved short-to-medium-range ensemble precipitation guidance relative to GEFS, while extreme precipitation intensity and ensemble calibration remain areas for further improvement.
\end{abstract}

\section{Introduction}
\noindent 
Accurate precipitation prediction is essential for disaster mitigation, water resource management, and sustainable development. Over the last decade, improvements in high-performance computing have greatly advanced numerical weather prediction (NWP). Traditional NWP systems rely on explicitly simulating atmospheric processes by solving large sets of partial differential equations (PDEs) that govern fluid dynamics and thermodynamics \cite{ref-book1}. While physically rigorous, this simulation-based approach is computationally demanding and often slow, as it requires massive resources to integrate the equations forward in time at high resolution. Traditional deterministic NWP systems generate a single forecast trajectory from given initial conditions. While such forecasts can be accurate in the short range, they fail to capture the inherent uncertainty of the atmosphere. This limitation motivates the use of ensemble prediction systems, in which multiple forecasts are produced by perturbing the initial conditions and integrating each perturbed state forward in time \cite{ref-book2}. For example, the state-of-the-art European Centre for Medium-Range Weather Forecasts (ECMWF) ensemble (ENS) consists of one control forecast and 50 perturbed members, providing medium-range predictions up to 15 days ahead. In practice, ensemble forecasts are essential because a single deterministic forecast can be misleading: it does not convey the range of possible outcomes. By contrast, ensembles quantify uncertainty by showing the spread of scenarios, which is critical for decision-making in sectors such as agriculture and disaster risk management. A reliable ensemble not only indicates the likelihood of specific events—for instance, a 70\% chance of exceeding a temperature threshold—but also ensures that such probabilities align with observed frequencies \cite{ref-book3}, thereby providing both sharpness and reliability in forecasts.

Recent advances in machine learning (ML) have opened new possibilities for weather forecasting, providing substantially faster and precise predictions compared to traditional physics-based NWP systems \cite{ref-book4}. For example, recent approaches such as FourCastNet have demonstrated dramatic computational advantages. In producing a 100-member, 24-hour ensemble forecast, FourCastNet is approximately 145,000 times faster than the ECMWF Integrated Forecasting System (IFS) at 30 km resolution, and an estimated 45,000 times faster at 18 km resolution, while also consuming substantially less energy \cite{ref-book5}. Models such as Gencast and FuXi-ENS have shown that ML-based systems can surpass state-of-the-art NWP ensembles in medium-range forecasts \cite{ref-book6, ref-book7}, highlighting a paradigm shift in weather prediction. These advances are driven not only by architectural innovations, such as Transformers and diffusion models, but also by the availability of high-quality, large-scale historical weather datasets such as fifth-generation ECMWF reanalysis (ERA5) \cite{ref-book8}.

More importantly, traditional NWP models exhibit systematic biases in precipitation representation, with rainfall simulated to occur too frequently and at intensities that are too weak, a deficiency that has been consistently reported in intercomparison studies \cite{ref-book22}. In fact, precipitation is among the most difficult atmospheric variables to predict, as it arises from highly nonlinear and multiscale processes, and its predictability is far lower than that of smoother variables such as temperature or pressure, since it strongly depends on convection and localized processes and requires the simultaneous handling of initial condition errors, multiscale interactions, and rapidly evolving convective systems \cite{ref-book22, ref-book19, ref-book21}. Despite the remarkable success of machine learning-based weather forecasting in recent years, current ML models remain less effective for precipitation, and precipitation forecasts still face significant challenges. First, uncertainties in the initial conditions and observational datasets propagate and grow rapidly during model integration \cite{ref-book19}. Second, the reliability of precipitation datasets is primarily constrained by the number and spatial coverage of ground stations, the accuracy of satellite retrieval algorithms and the limitations of data assimilation models \cite{ref-book21}. While ERA5 provides a comprehensive and dynamically consistent set of atmospheric variables, its precipitation field is model-derived and can differ substantially from observational precipitation products, especially for localized and intense rainfall (see Appendix~\ref{sec:era5_appendix} for a sensitivity comparison using different precipitation targets). For this reason, several recent ML weather forecasting studies have treated precipitation verification separately from standard ERA5-based evaluation protocols \cite{ref-book20}. In this study, we therefore use ERA5 for atmospheric predictors and Integrated Multi-satellite Retrievals for Global Precipitation Measurement (IMERG) as the precipitation training target and verification reference. 

In this paper, we introduce Colorado State University Precipitation foreCAST (CSU-PCAST) framework, a deep learning-based medium-range ensemble weather forecasting model designed to improve precipitation forecasts relative to operational ensemble baselines at a fine spatial resolution of \(0.25^\circ\). The model produces 15-day forecasts every 6 hours, conditioned on 6 atmospheric variables at 8 pressure levels and 9 surface variables, with an emphasis on precipitation prediction. To enhance precipitation forecasting, we adopt the global IMERG as a precipitation label. Specifically, training is performed on 21 years (1998-2018) of ERA5 reanalysis and IMERG precipitation data at 0.25° resolution, where ERA5 provides 57 variables (6×8+9) and IMERG serves as the precipitation reference. The almost fair Continuous Ranked Probability Score (afCRPS) is used for optimization to better capture ensemble uncertainty in precipitation. Unlike SEEDS \cite{li2024generative}, GenCast, and FuXi-ENS, our model is evaluated not on ERA5 reanalysis but against operational forecasts, aligning the assessment with real-world forecasting practice. Furthermore, instead of relying on the diffusion process, ensemble data assimilation perturbations, or operational ensemble members, our approach represents uncertainty by directly embedding noise into the Transformer blocks.

\section{Methods}

\subsection{CSU-PCAST model description}
\noindent CSU-PCAST builds on the FuXi-style autoregressive architecture, which uses Swin Transformer blocks to model spatiotemporal dependencies in atmospheric fields \cite{ref-book13}. Unlike FuXi, CSU-PCAST introduces stochastic perturbations into the high-dimensional latent representation to generate ensemble diversity. The decoder also uses two output branches: one branch predicts the atmospheric and surface state variables that are used as inputs in subsequent autoregressive steps, while the other predicts 6-h total precipitation as a diagnostic output. In this setup, precipitation is not fed back as an input variable, but is predicted separately from the evolving atmospheric state.

\begin{figure*}[htbp]
\centerline{\includegraphics[width=1\textwidth]{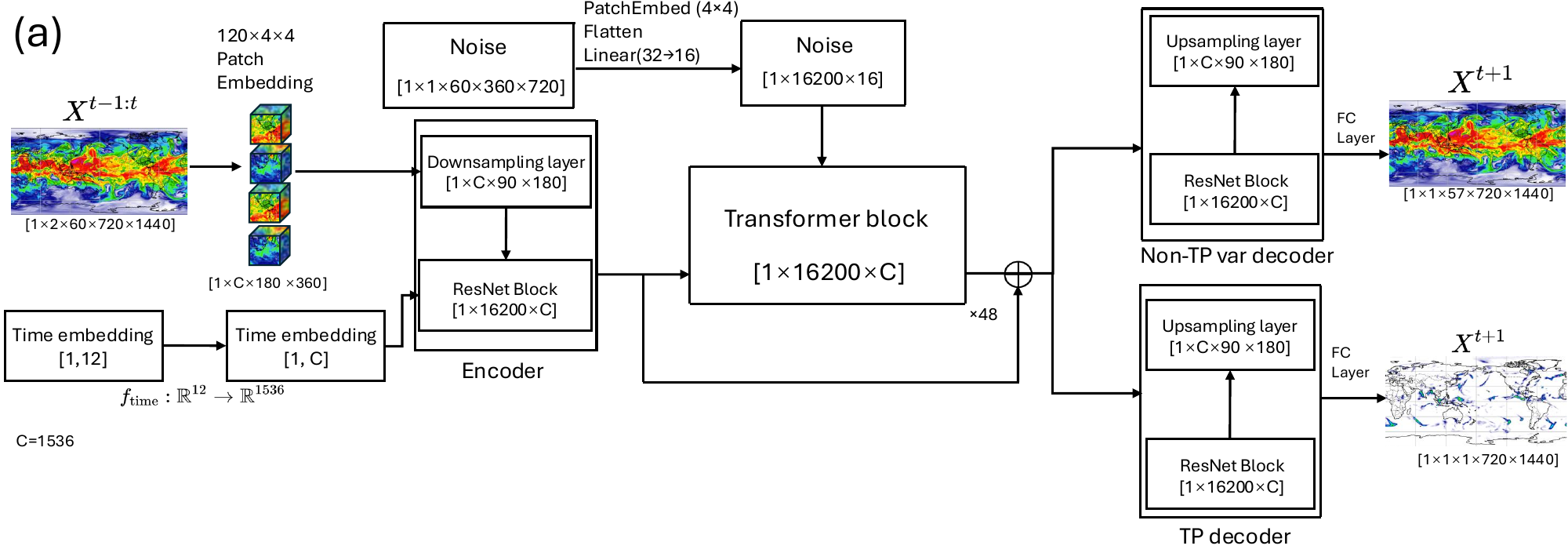}}
    \centerline{\includegraphics[width=1\textwidth]{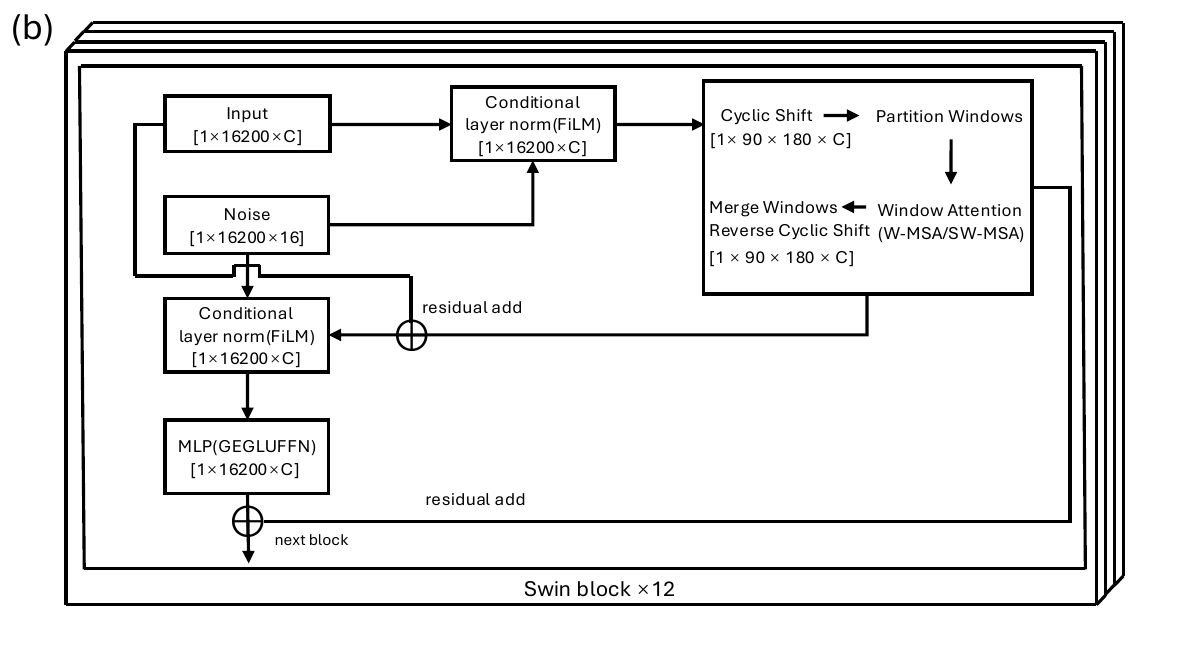}}
    \caption{Architecture of CSU-PCAST. The panel (a) illustrates the overall model structure, including patch embedding, the U-Transformer backbone, and fully connected output layers. The panel (b) details the Transformer block within the U-Transformer backbone, where each Transformer block contains four Swin layers and each Swin layer consists of 12 Swin blocks.}
\label{fig:model_structure}
\end{figure*}

The model consists of four main components: patch embedding, a U-Transformer backbone, a noise-conditioning module, and fully connected output layers, as shown in Fig.~\ref{fig:model_structure}. For compatibility with the patch-embedding and encoder--decoder structure, the standard \(721 \times 1440\) input fields are interpolated to the model's internal \(720 \times 1440\) grid. The meteorological input tensor has shape \(1 \times 2 \times 57 \times 720 \times 1440\), corresponding to batch size, two preceding time steps \((t-1,t)\), atmospheric and surface variables, latitude, and longitude. Fig.~\ref{fig:model_structure} shows only the 57 atmospheric and surface variables. In the actual model input, we also include the three geographical variables listed in Table~\ref{tab:variables}, giving 60 input channels in total. \footnote{The released inference checkpoint accepts a 63-channel input tensor. The final three channels are constant-one reserved channels introduced during model design to support flexible fine-tuning. They do not carry additional meteorological information and are fixed during inference, so the effective meteorological input remains 60 channels.}

The model first concatenates atmospheric inputs from two consecutive time steps and passes them through a space-time patch embedding module. This module merges the temporal and channel dimensions and reduces the spatial resolution through patch embedding. Temporal information is represented using sinusoidal encodings of day-of-year and hour-of-day, followed by sine and cosine transformations and projection into the latent space. These time embeddings are added to the embedded features. Stochastic perturbations are then injected into the latent representation through Gaussian noise conditioning. The resulting features are processed by a U-Transformer encoder--decoder backbone, which captures multi-scale spatial and temporal structure. The final outputs are reconstructed through upsampling and fully connected layers on the model's internal $720 \times 1440$ computational grid. 

\subsubsection{Patch embedding}
\noindent To reduce computational cost, CSU-PCAST uses a two-dimensional convolutional patch embedding module. The temporal and variable dimensions are first merged into the channel dimension, converting the input from $[1,2,60,720,1440]$ to $[1,120,720,1440]$. A 2D convolution with kernel size and stride $4 \times 4$ is then applied to form non-overlapping spatial patches. The resulting feature map has shape $[1,C,180,360]$, where the number of output channels is set to \(C=1536\).

\subsubsection{U-Transformer}
\noindent The U-Transformer is the backbone of CSU-PCAST. It contains a convolutional encoder, a stack of Swin Transformer V2 blocks \cite{liu2022swin}, and a dual-branch decoder. The input fields are first mapped to patch-level features using the $4 \times 4$ convolutional patch embedding. The encoder then processes these features hierarchically. Each encoder stage includes a downsampling layer followed by residual blocks. The downsampling layer uses a strided $3 \times 3$ convolution with stride 2 to halve the spatial resolution. The residual blocks refine the downsampled features using Group Normalization, SiLU activation, and $3 \times 3$ convolutions with skip connections. Within these residual blocks, convolutions use periodic padding in the longitudinal direction and zero padding in the latitudinal direction, preserving continuity across the global longitude boundary.

Temporal information is injected through time embeddings. A two-layer multilayer perceptron (MLP) projects the time embedding, which is then used in the residual blocks through Feature-wise Linear Modulation (FiLM) \cite{perez2018film}. The projected embedding is split into scale and shift terms that modulate the normalized feature maps through a learned affine transformation. This conditioning allows the encoder and decoder to adapt their features to the forecast lead time.

The transformer module contains four Swin layers, each with 12 Swin Transformer V2 blocks, for a total of 48 Swin blocks. Within each Swin layer, regular window attention and shifted-window attention are applied alternately. This structure keeps attention computationally efficient while allowing information exchange across neighboring windows. The attention operation follows the cosine attention formulation used in Swin Transformer V2 \cite{liu2022swin}:
\begin{equation}
\text{Attention}(\mathbf{Q}, \mathbf{K}, \mathbf{V}) =
\text{Softmax}\left(s \cdot
\cos(\mathbf{Q}, \mathbf{K}) + \mathbf{B}\right)\mathbf{V},
\end{equation}
where $\mathbf{Q}$, $\mathbf{K}$, and $\mathbf{V}$ are the query, key, and value matrices, \(s\) is a learnable logit scale, and \(\mathbf{B}\) is the continuous relative position bias generated by a small MLP from relative spatial coordinates. Each Swin block also includes a GEGLU-based \cite{shazeer2020glu} feed-forward network after the attention module.

To introduce stochastic conditioning, CSU-PCAST replaces standard Layer Normalization in the Swin blocks with Conditional Layer Normalization (CLN). A low-resolution noise tensor is provided as an additional input. In the default configuration, this tensor has half the spatial resolution of the original input and is embedded using a separate patch-embedding module with the same $4 \times 4$ patch size. This produces token-level noise features aligned with the transformer features after encoder downsampling. CLN uses the noise embedding to generate dynamic scale and bias parameters that modulate the normalized features channel by channel:
\begin{equation}
\text{CLN}(\mathbf{x}, \mathbf{z}) =
\gamma(\mathbf{z}) \cdot \text{LN}(\mathbf{x}) + \beta(\mathbf{z}),
\end{equation}
where \(\mathbf{z}\) is the embedded noise condition, and \(\gamma(\cdot)\) and \(\beta(\cdot)\) are learned linear projections. CLN is applied before both the attention and feed-forward sublayers in every Swin block, allowing the injected noise to affect the latent representation throughout the transformer stack.

The decoder uses separate branches for precipitation and non-precipitation variables. After the Swin Transformer stack, the outputs from the four Swin layers are normalized, concatenated along the channel dimension, and compressed using branch-specific fusion layers. One fusion layer is used for precipitation, and the other is used for the remaining atmospheric variables. The fused features are reshaped into spatial feature maps and passed through two symmetric upsampling pathways. Each pathway upsamples the features using transposed convolutions and merges them with the corresponding encoder features through skip connections. The decoder residual blocks also use FiLM-based temporal conditioning so that the reconstructed fields remain tied to the target forecast lead time.

The non-precipitation branch predicts 57 atmospheric variables through a patch-level linear projection, while the precipitation branch predicts one total-precipitation channel. The patch-level outputs are reshaped back into spatial fields. If needed, bilinear interpolation is applied to recover the model's internal $720 \times 1440$ resolution. The two branch outputs are concatenated to form the final 58-channel forecast. The final forecasts are mapped back to the standard \(0.25^\circ\), \(721 \times 1440\) grid before verification.
\subsection{Model training and fine-tuning}

\noindent The model is trained under an autoregressive forecasting paradigm. Given two consecutive input states, the network predicts the atmospheric state at the next forecast lead time; during multi-step training, each prediction is fed back into the input sequence to generate the following step. Ensemble forecasts are generated by repeated stochastic forward passes: for each forecast step, we run the network $M$ times with independently sampled noise inputs while keeping the meteorological input state fixed, producing $M$ ensemble members. During training, we use $M=2$. In multi-step rollouts, each member is then advanced autoregressively using its own previous prediction, so the ensemble trajectories evolve independently.

We use the same ensemble loss for two target modes, but apply it to different output channels: the first 57 channels for non-precipitation training and the final precipitation channel for precipitation fine-tuning. The training is organized into three main stages: one-step non-precipitation pre-training, multi-step non-precipitation fine-tuning, and one-step precipitation training with the main network frozen. A separate Global Forecast System (GFS) fine-tuning experiment was conducted for diagnostic purposes, but it was not included in the final CSU-PCAST configuration.

For a rollout with $S$ forecast steps, the objective is
\begin{equation}
\mathcal{L}_{S}
=
\mathcal{L}_{\mathrm{wafCRPS}}
\left(
\left\{
\hat{\mathbf{X}}^{(m),t+S}_{\mathcal{C}}
\right\}_{m=1}^{M},
\mathbf{X}^{t+S}_{\mathcal{C}}
\right),
\label{eq:training_objective}
\end{equation}
where $\{\hat{\mathbf{X}}^{(m),t+S}_{\mathcal{C}}\}_{m=1}^{M}$ denotes the $M$-member ensemble forecast for the selected target
channels at the final unrolled lead step, $\mathbf{X}^{t+S}_{\mathcal{C}}$ is the corresponding target, and $\mathcal{C}$ denotes the set of channels used in the loss. During non-precipitation training, $\mathcal{C}$ contains the 57 prognostic atmospheric and surface channels, whereas during precipitation fine-tuning, $\mathcal{C}$ contains only the precipitation channel. For multi-step fine-tuning, the model is unrolled autoregressively for $S$ steps. Intermediate predictions are fed back into the input sequence with stopped gradients, so they determine the forecast state seen at later lead times but do not directly contribute gradients. The optimization objective is therefore applied only at the final unrolled lead step.

We optimize the weighted almost-fair Continuous Ranked Probability Score (wafCRPS), following the almost-fair CRPS (afCRPS) introduced for AIFS-CRPS~\cite{lang2026aifs} and its multi-scale extension~\cite{ec_afcrps}. This loss builds on finite-ensemble corrections to the CRPS developed for fair ensemble verification~\cite{ferro2013fair,leutbecher2019ensemble}. The ensemble-spread term is computed only over distinct member pairs, avoiding self-pair contributions of the form $|\hat{x}_j-\hat{x}_j|=0$ that would otherwise bias the estimated spread when $M$ is small:
\begin{equation}
\mathcal{L}_{\mathrm{wafCRPS}}
=
\frac{1}{BCHW}
\sum_{i=1}^{B}
\sum_{c=1}^{C}
\sum_{h=1}^{H}
\sum_{w=1}^{W}
\omega_{\mathrm{var}}(c)
\omega_{\mathrm{lat}}(h)
\,
\mathrm{afCRPS}
\left(
\{\hat{x}_{i,m,c,h,w}\}_{m=1}^{M},
x_{i,c,h,w}
\right),
\label{eq:wafcrps_loss}
\end{equation}
where $B$ is the batch size, $C$ is the number of selected target channels, $H$ and $W$ are the spatial dimensions, and $M$ is the ensemble size. For each grid point and variable, the almost-fair CRPS is
\begin{equation}
\mathrm{afCRPS}
=
\frac{1}{M}\sum_{m=1}^{M}|\hat{x}_m-x|
-
\frac{1-\epsilon}{2M(M-1)}
\sum_{\substack{j,k=1 \\ j\ne k}}^{M}
|\hat{x}_j-\hat{x}_k|,
\qquad
\epsilon=\frac{1-\alpha}{M},\quad \alpha=0.95.
\label{eq:afcrps}
\end{equation}

The latitude-dependent weight accounts for the unequal physical area represented by grid cells on a latitude--longitude grid. Grid cells at high latitudes cover smaller physical areas than those near the equator; without area weighting, high-latitude regions would contribute disproportionately to the spatially averaged loss \cite{ref-book20,graphcast}. We therefore adopt a cosine-latitude weighting scheme,
\begin{equation*}
\omega_{\mathrm{lat}}(h)=
\frac{\cos(\phi_h)}
{H^{-1}\sum_{h=1}^{H}\cos(\phi_h)},
\end{equation*}
where $\phi_h$ is the latitude of row $h$. The normalization gives the latitude weights unit mean, preserving the overall scale of the loss.

The variable-dependent weight depends on the selected training mode. For precipitation-only fine-tuning, the loss reduces to the latitude-weighted afCRPS over the single precipitation channel, and $\omega_{\mathrm{var}}(c)=1$. For non-precipitation training, the 57 channels consist of six upper-air variables on eight pressure levels and nine surface variables. We apply a vertical weighting scheme only within the upper-air fields \cite{lang2026aifs, graphcast}:
\begin{equation*}
p_\ell\in\{200,250,300,400,500,600,700,850\}\ \mathrm{hPa},\qquad
\tilde{\omega}_{\ell}=\max\left(\frac{p_\ell}{\overline{p}},\tau\right),\quad
\tau=0.2,
\end{equation*}
\begin{equation*}
\omega_\ell=
\frac{\tilde{\omega}_{\ell}}{L^{-1}\sum_{\ell=1}^{L}\tilde{\omega}_{\ell}},
\qquad
\omega_{\mathrm{surface}}=1.
\end{equation*}
The pressure-level weights are normalized to have unit mean across the selected levels and are repeated for all six upper-air variables. This redistributes loss within each upper-air variable, shifting emphasis toward lower-tropospheric levels while preserving the total weight of each variable. Surface variables are assigned a neutral per-channel weight of one, so the relative contribution of upper-air and surface variables remains primarily determined by the number of channels in each group.

During autoregressive rollout, each ensemble member is updated independently. After each forecast step, the predicted non-precipitation variables are concatenated with the static masks and used, together with the previous input state, as the input for the next lead time. This enables the model to learn forecast trajectories over multiple lead times while retaining separate optimization targets for precipitation and non-precipitation variables.

\subsubsection{Non-precipitation pre-training}
\noindent In the pre-training stage, the model is trained to predict $\mathbf{X}^{t+1}$ from the two preceding time steps $\mathbf{X}^{t-1}$ and $\mathbf{X}^{t}$ for non-precipitation variables only. The precipitation decoder remains frozen during this stage, and only parameters associated with non-precipitation variables are updated.

Training is conducted with an ensemble size of $M=2$. We train the model on 32 NVIDIA A100 GPUs for approximately 27{,}000 iterations (about 30 hours). The AdamW optimizer is adopted with $\beta_1=0.9$, $\beta_2=0.95$, weight decay of 0.01, and a learning rate decayed from $3\times10^{-4}$ to $1\times10^{-5}$, following the configuration of FuXi \cite{ref-book13}.

Due to the large model size (approximately 1.5 billion parameters), Fully Sharded Data Parallel (FSDP) training with bfloat16 precision and a hybrid sharding strategy is employed to mitigate memory constraints. This stage focuses on stabilizing large-scale atmospheric and surface representations before incorporating precipitation forecasting.

\subsubsection{Non-precipitation multi-step fine-tuning}
\noindent In the multi-step fine-tuning stage, the model is fine-tuned autoregressively to improve stability over extended forecast horizons. All configurations remain identical to pre-training, except that the learning rate is further reduced to $1\times10^{-7}$. During fine-tuning, the model is unrolled for up to 8 autoregressive forecast steps. At each step, the model takes as input the two most recent time frames of non-precipitation variables (57 channels) and predicts the full set of outputs (57 non-precipitation variables plus 1 precipitation variable). The 57 non-precipitation outputs are extracted and fed back into the next step, replacing the oldest input frame, while the precipitation output is discarded since its decoder remains frozen in this stage. This iterative training process enables the model to adapt to error accumulation and learn temporal consistency across longer forecast lead times.

\subsubsection{One-step precipitation training}
\noindent In the precipitation training stage, the precipitation decoder is trained while all other model parameters remain frozen. Unlike non-precipitation variables, which require multi-step fine-tuning to mitigate autoregressive error accumulation, precipitation is trained only for single-step prediction. During inference, forecasts are generated autoregressively by feeding predicted states back into the model.

Since most parameters are frozen in this stage, we switch from FSDP to Distributed Data Parallel (DDP) for improved computational efficiency.

\subsubsection{Fine-tuning on GFS}
\noindent We conducted an additional experiment to fine-tune the model with GFS analyses after the multi-step fine-tuning stage and before the final one-step precipitation fine-tuning stage. In this experiment, GFS analysis fields were used in place of ERA5 reanalysis fields as the model inputs. This step was performed with a very small learning rate ($1\times10^{-7}$), with the goal of adapting the model to the input distribution encountered during operational inference, where forecasts are initialized from GFS analyses. 

However, this additional fine-tuning did not improve the forecasts. Instead, the GFS-fine-tuned model performed slightly worse than the checkpoint trained only with ERA5-based inputs. The Critical Success Index (CSI) scores at light precipitation thresholds (0.1--5 mm) remained broadly comparable, whereas the scores at heavier precipitation thresholds (10--20 mm) degraded more noticeably and decreased more rapidly toward zero with lead time. One possible explanation is that the extra fine-tuning step may still introduce small changes to the learned parameters, potentially causing the model to partially absorb errors or biases in the GFS analyses rather than correcting them. It may also slightly weaken the large-scale structures learned during the non-precipitation variables of ERA5 pretraining. Therefore, we do not attribute the degradation to a single mechanism, but our results suggest that direct fine-tuning on GFS analyses is not necessarily beneficial for heavy-precipitation prediction in the current setting.

\subsection{Evaluation methods}
\noindent The evaluation is conducted using deterministic metrics, probabilistic metrics, and extreme precipitation event analysis, following standard forecast-verification practice \cite{jolliffe2012forecast}. Specifically, the deterministic evaluation includes latitude-weighted root-mean-square error (RMSE) and root-mean-square error Skill Score (RMSESS) for the ensemble-mean forecast, as well as member-level CSI, CSI Skill Score (CSISS), and ensemble-pooled precipitation frequency bias, which are calculated as follows:
\begin{equation}
\begin{gathered}
\begin{aligned}
\mathrm{RMSE}(c,\tau)
&=
\frac{1}{|D_{\mathrm{eval}}|}
\sum_{t_0 \in D_{\mathrm{eval}}}
\sqrt{
  \frac{1}{N}
  \sum_{i \in \Omega}
  a_i
  \left(
    \hat{X}_{c,i}^{t_0+\tau}
    -
    X_{c,i}^{t_0+\tau}
  \right)^2
},
\qquad
a_i
=
\frac{\cos(\phi_i)}
{
  \frac{1}{N_{\mathrm{lat}}}
  \sum_{k=1}^{N_{\mathrm{lat}}}
  \cos(\phi_k)
},
\end{aligned}
\\[0.5em]
\begin{aligned}
\mathrm{RMSESS}(c,\tau)
&=
\frac{
  \mathrm{RMSE}_{\mathrm{GEFS}}(c,\tau)
  -
  \mathrm{RMSE}_{\mathrm{PCAST}}(c,\tau)
}{
  \mathrm{RMSE}_{\mathrm{GEFS}}(c,\tau)
}
=
1
-
\frac{
  \mathrm{RMSE}_{\mathrm{PCAST}}(c,\tau)
}{
  \mathrm{RMSE}_{\mathrm{GEFS}}(c,\tau)
}.
\end{aligned}
\end{gathered}
\end{equation}

where \(t_0\) denotes the forecast initialization time, \(\tau\) denotes the forecast lead time, \(c\) denotes the variable or channel being evaluated, and \(D_{\mathrm{eval}}\) is the set of evaluation initialization times. \(\Omega\) denotes the set of valid spatial grid points, with \(N = |\Omega|\). \(\hat{X}_{c,i}^{t_0+\tau}\) and \(X_{c,i}^{t_0+\tau}\) are the forecast and reference values, respectively, for variable \(c\) at grid point \(i\) and valid time \(t_0+\tau\). The latitude weight \(a_i\) is defined from the latitude\(\phi_i\) of grid point \(i\), normalized by the mean cosine latitude over all \(N_{\mathrm{lat}}\) latitude rows. The RMSESS uses Global Ensemble Forecast System (GEFS) as the reference forecast. Positive values indicate that PCAST has lower RMSE than GEFS, zero indicates equal RMSE, and negative values indicate higher RMSE for PCAST.

\begin{equation}
\begin{gathered}
\begin{aligned}
\mathrm{CSI}(q,\tau)
&=
\frac{1}{|D_{\mathrm{eval}}|}
\sum_{t_0 \in D_{\mathrm{eval}}}
\frac{1}{|\mathcal{M}|}
\sum_{r \in \mathcal{M}}
\frac{
  H_{r}^{t_0,\tau}(q)
}{
  H_{r}^{t_0,\tau}(q)
  +
  M_{r}^{t_0,\tau}(q)
  +
  F_{r}^{t_0,\tau}(q)
},
\end{aligned}
\\[0.5em]
\begin{aligned}
H_{r}^{t_0,\tau}(q)
&=
\sum_{i \in \Omega}
\mathbf{1}
\left[
  \hat{P}_{r,i}^{t_0+\tau} > q
  \ \mathrm{and}\
  P_i^{t_0+\tau} > q
\right],
\end{aligned}
\\[0.5em]
\begin{aligned}
M_{r}^{t_0,\tau}(q)
&=
\sum_{i \in \Omega}
\mathbf{1}
\left[
  \hat{P}_{r,i}^{t_0+\tau} \le q
  \ \mathrm{and}\
  P_i^{t_0+\tau} > q
\right],
\end{aligned}
\\[0.5em]
\begin{aligned}
F_{r}^{t_0,\tau}(q)
&=
\sum_{i \in \Omega}
\mathbf{1}
\left[
  \hat{P}_{r,i}^{t_0+\tau} > q
  \ \mathrm{and}\
  P_i^{t_0+\tau} \le q
\right],
\end{aligned}
\\[0.5em]
\begin{aligned}
\mathrm{CSISS}(q,\tau)
&=
\frac{
  \mathrm{CSI}_{\mathrm{PCAST}}(q,\tau)
  -
  \mathrm{CSI}_{\mathrm{GEFS}}(q,\tau)
}{
  1
  -
  \mathrm{CSI}_{\mathrm{GEFS}}(q,\tau)
}.
\end{aligned}
\end{gathered}
\end{equation}

where \(q\) denotes the precipitation threshold, \(r\) indexes an individual ensemble member, and \(\mathcal{M}\) is the set of ensemble members. The member-level CSI is computed by evaluating each ensemble member independently against the reference precipitation field, followed by averaging over members and initialization times. \(H\), \(M\), and \(F\) denote hits, misses, and false alarms, corresponding to true positives, false negatives, and false positives, respectively. The CSISS uses GEFS as the reference forecast; positive values indicate higher CSI for PCAST than GEFS.

\begin{equation}
\begin{gathered}
\begin{aligned}
\mathrm{Bias}(q,\tau)
&=
\frac{
  \sum_{t_0 \in D_{\mathrm{eval}}}
  \sum_{r \in \mathcal{M}}
  \sum_{i \in \Omega}
  \mathbf{1}
  \left[
    \hat{P}_{r,i}^{t_0+\tau} > q
  \right]
}{
  \sum_{t_0 \in D_{\mathrm{eval}}}
  \sum_{r \in \mathcal{M}}
  \sum_{i \in \Omega}
  \mathbf{1}
  \left[
    P_i^{t_0+\tau} > q
  \right]
}.
\end{aligned}
\end{gathered}
\end{equation}

where \(\hat{P}_{r,i}^{t_0+\tau}\) denotes the forecast precipitation from ensemble member \(r\) at grid point \(i\) and valid time \(t_0+\tau\), while \(P_i^{t_0+\tau}\) denotes the corresponding IMERG reference precipitation. The indicator function counts threshold-exceeding precipitation events. A bias value greater than one indicates overforecasting of event frequency, while a value less than one indicates underforecasting. 

In addition, we evaluate the quality of the ensemble using the following probabilistic diagnostics: latitude-weighted CRPS \cite{hersbach2000decomposition}, Brier Skill Score \cite{brier1950verification}, Rank Histogram \cite{hamill2001interpretation}, Quantile-Quantile (Q-Q) plot, and spread-to-RMSE ratio, for quantifying the quality of probabilistic precipitation forecasts.

\begin{equation}
\begin{gathered}
\begin{aligned}
\mathrm{CRPS}(c,\tau)
&=
\frac{1}{|D_{\mathrm{eval}}|}
\sum_{t_0 \in D_{\mathrm{eval}}}
\frac{1}{N}
\sum_{i \in \Omega}
a_i
\left[
  \frac{1}{|\mathcal{M}|}
  \sum_{r \in \mathcal{M}}
  \left|
    \hat{X}_{c,r,i}^{t_0+\tau}
    -
    X_{c,i}^{t_0+\tau}
  \right|
  -
  \frac{1}{2|\mathcal{M}|^2}
  \sum_{r \in \mathcal{M}}
  \sum_{s \in \mathcal{M}}
  \left|
    \hat{X}_{c,r,i}^{t_0+\tau}
    -
    \hat{X}_{c,s,i}^{t_0+\tau}
  \right|
\right].
\end{aligned}
\end{gathered}
\end{equation}

where \(\hat{X}_{c,r,i}^{t_0+\tau}\) denotes the forecast value from ensemble member \(r\). The first term measures the ensemble error relative to the reference, while the second term accounts for ensemble spread. Lower CRPS indicates a better calibrated and more accurate ensemble forecast.

\begin{equation}
\begin{gathered}
\begin{aligned}
p_i^{t_0,\tau}(q)
&=
\frac{1}{|\mathcal{M}|}
\sum_{r \in \mathcal{M}}
\mathbf{1}
\left[
  \hat{P}_{r,i}^{t_0+\tau} \ge q
\right],
\qquad
o_i^{t_0,\tau}(q)
=
\mathbf{1}
\left[
  P_i^{t_0+\tau} \ge q
\right],
\end{aligned}
\\[0.5em]
\begin{aligned}
\mathrm{BS}(q,\tau)
&=
\frac{1}{|D_{\mathrm{eval}}|}
\sum_{t_0 \in D_{\mathrm{eval}}}
\frac{1}{N}
\sum_{i \in \Omega}
a_i
\left(
  p_i^{t_0,\tau}(q)
  -
  o_i^{t_0,\tau}(q)
\right)^2,
\end{aligned}
\\[0.5em]
\begin{aligned}
\mathrm{BSS}(q,\tau)
&=
\frac{1}{|D_{\mathrm{eval}}|}
\sum_{t_0 \in D_{\mathrm{eval}}}
\left[
  1
  -
  \frac{
    \mathrm{BS}_{\mathrm{PCAST}}^{t_0}(q,\tau)
  }{
    \mathrm{BS}_{\mathrm{GEFS}}^{t_0}(q,\tau)
  }
\right].
\end{aligned}
\end{gathered}
\end{equation}

where \(p_i^{t_0,\tau}(q)\) is the ensemble probability of exceeding threshold \(q\), and \(o_i^{t_0,\tau}(q)\) is the binary observed event indicator. GEFS is used as the reference forecast in the Brier Skill Score (BSS), so positive BSS values indicate that PCAST has lower Brier Score (BS) than GEFS.

\begin{equation}
\begin{gathered}
\begin{aligned}
B_i^{t_0,\tau}
&=
\sum_{r \in \mathcal{M}}
\mathbf{1}
\left[
  \hat{P}_{r,i}^{t_0+\tau}
  <
  P_i^{t_0+\tau}
\right],
\end{aligned}
\\[0.5em]
\begin{aligned}
E_i^{t_0,\tau}
&=
\sum_{r \in \mathcal{M}}
\mathbf{1}
\left[
  \hat{P}_{r,i}^{t_0+\tau}
  =
  P_i^{t_0+\tau}
\right],
\end{aligned}
\\[0.5em]
\begin{aligned}
R_i^{t_0,\tau}
&=
B_i^{t_0,\tau}
+
U_i^{t_0,\tau},
\qquad
U_i^{t_0,\tau}
\sim
\mathrm{Uniform}
\left\{
  0,\ldots,E_i^{t_0,\tau}
\right\},
\end{aligned}
\\[0.5em]
\begin{aligned}
C_k
&=
\sum_{t_0 \in D_{\mathrm{eval}}}
\sum_{\tau \in \mathcal{L}}
\sum_{i \in \Omega}
\mathbf{1}
\left[
  R_i^{t_0,\tau} = k
\right],
\qquad
k = 0,\ldots,|\mathcal{M}|,
\end{aligned}
\\[0.5em]
\begin{aligned}
\widetilde{C}_k
&=
\frac{
  C_k
}{
  \frac{1}{|\mathcal{M}|+1}
  \sum_{\ell=0}^{|\mathcal{M}|} C_\ell
}.
\rho_k = \frac{k}{|\mathcal{M}|}, \qquad k=0,\ldots,|\mathcal{M}|.
\end{aligned}
\end{gathered}
\end{equation}

where \(\mathcal{L}\) denotes the set of lead times included in the diagnostic aggregation. \(R_i^{t_0,\tau}\) is the randomized rank of the observation among the ensemble members, \(C_k\) is the count in rank bin \(k\), and \(\widetilde{C}_k\) is the count normalized by the expected count under a uniform rank histogram. For plotting, the rank bin is normalized as \(\rho_k=k/|\mathcal{M}|\), so that the rank histograms are displayed on the unit interval \([0,1]\).
\begin{equation}
\begin{gathered}
\begin{aligned}
\widehat{F}_{\mathrm{fcst}}(x)
&=
\frac{
  1
}{
  |D_{\mathrm{eval}}|\,|\mathcal{L}|\,|\mathcal{M}|\,N
}
\sum_{t_0 \in D_{\mathrm{eval}}}
\sum_{\tau \in \mathcal{L}}
\sum_{r \in \mathcal{M}}
\sum_{i \in \Omega}
\mathbf{1}
\left[
  \hat{P}_{r,i}^{t_0+\tau} \le x
\right],
\end{aligned}
\\[0.5em]
\begin{aligned}
\widehat{F}_{\mathrm{obs}}(x)
&=
\frac{
  1
}{
  |D_{\mathrm{eval}}|\,|\mathcal{L}|\,N
}
\sum_{t_0 \in D_{\mathrm{eval}}}
\sum_{\tau \in \mathcal{L}}
\sum_{i \in \Omega}
\mathbf{1}
\left[
  P_i^{t_0+\tau} \le x
\right],
\end{aligned}
\\[0.5em]
\begin{aligned}
Q_{\mathrm{fcst}}(p)
&=
\inf
\left\{
  x:
  \widehat{F}_{\mathrm{fcst}}(x) \ge p
\right\},
\qquad
Q_{\mathrm{obs}}(p)
=
\inf
\left\{
  x:
  \widehat{F}_{\mathrm{obs}}(x) \ge p
\right\}.
\end{aligned}
\end{gathered}
\end{equation}

where \(\widehat{F}_{\mathrm{fcst}}\) and \(\widehat{F}_{\mathrm{obs}}\) are the empirical marginal cumulative distribution functions of ensemble forecast precipitation and IMERG precipitation, respectively. \(Q_{\mathrm{fcst}}(p)\) and \(Q_{\mathrm{obs}}(p)\) are the corresponding empirical quantile functions. The Q-Q plot places \(Q_{\mathrm{obs}}(p)\) on the x-axis and \(Q_{\mathrm{fcst}}(p)\) on the y-axis.

\begin{equation}
\begin{gathered}
\begin{aligned}
\overline{X}_{c,i}^{t_0+\tau}
&=
\frac{1}{|\mathcal{M}|}
\sum_{r \in \mathcal{M}}
\hat{X}_{c,r,i}^{t_0+\tau},
\end{aligned}
\\[0.5em]
\begin{aligned}
S^{t_0}(c,\tau)
&=
\sqrt{
  \frac{1}{N}
  \sum_{i \in \Omega}
  a_i
  \left[
    \frac{1}{|\mathcal{M}|-1}
    \sum_{r \in \mathcal{M}}
    \left(
      \hat{X}_{c,r,i}^{t_0+\tau}
      -
      \overline{X}_{c,i}^{t_0+\tau}
    \right)^2
  \right]
},
\end{aligned}
\\[0.5em]
\begin{aligned}
\mathrm{Spread/RMSE}(c,\tau)
&=
\frac{1}{|D_{\mathrm{eval}}|}
\sum_{t_0 \in D_{\mathrm{eval}}}
\frac{
  S^{t_0}(c,\tau)
}{
  \mathrm{RMSE}^{t_0}(c,\tau)
}.
\end{aligned}
\end{gathered}
\end{equation}

where \(\overline{X}_{c,i}^{t_0+\tau}\) is the ensemble mean and \(S^{t_0}(c,\tau)\) is the latitude-weighted ensemble spread for a single initialization time. A spread-to-RMSE ratio near one indicates that the ensemble spread is comparable to the ensemble-mean error.

For the extreme precipitation event analysis, we evaluate it using the following event-based diagnostic: 72-h accumulated precipitation over lead time 3-5, Fractions Skill Score (FSS) \cite{roberts2008scale}, spatial correlation, regional precipitation bias, ensemble exceedance probability, and regional BS. All metrics are computed over the regional domain covering the precipitation area. 

The 72-h accumulated precipitation is obtained by summing twelve consecutive 6-h
precipitation fields over lead days 3--5:
\begin{equation}
\begin{gathered}
\begin{aligned}
\hat{A}_{r,i}
&=
\sum_{\ell \in \mathcal{T}_{72}}
\hat{P}_{r,i}^{t_0+\ell},
\qquad
A_i
=
\sum_{\ell \in \mathcal{T}_{72}}
P_i^{t_0+\ell}.
\end{aligned}
\end{gathered}
\end{equation}

where \(\mathcal{T}_{72}\) denotes the twelve 6-h valid times included in the
72-h accumulation window, \(\hat{A}_{r,i}\) is the accumulated precipitation from
ensemble member \(r\), and \(A_i\) is the corresponding IMERG accumulation.

\begin{equation}
\begin{gathered}
\begin{aligned}
I_{r,i}(q)
&=
\mathbf{1}
\left[
  \hat{A}_{r,i} \ge q
\right],
\qquad
O_i(q)
=
\mathbf{1}
\left[
  A_i \ge q
\right],
\end{aligned}
\\[0.5em]
\begin{aligned}
f_{r,i}^{(w)}(q)
&=
\frac{
  \sum_{j \in \mathcal{N}_w(i)} I_{r,j}(q)
}{
  \sum_{j \in \mathcal{N}_w(i)} V_j
},
\qquad
g_i^{(w)}(q)
=
\frac{
  \sum_{j \in \mathcal{N}_w(i)} O_j(q)
}{
  \sum_{j \in \mathcal{N}_w(i)} V_j
},
\end{aligned}
\\[0.5em]
\begin{aligned}
\mathrm{FSS}_{r}(q,w)
&=
1
-
\frac{
  \frac{1}{|\Omega_w|}
  \sum_{i \in \Omega_w}
  \left(
    f_{r,i}^{(w)}(q)
    -
    g_i^{(w)}(q)
  \right)^2
}{
  \frac{1}{|\Omega_w|}
  \sum_{i \in \Omega_w}
  \left[
    \left(f_{r,i}^{(w)}(q)\right)^2
    +
    \left(g_i^{(w)}(q)\right)^2
  \right]
}.
\end{aligned}
\end{gathered}
\end{equation}

where \(q=50\) mm is the 72-h precipitation threshold, \(w\) is the neighborhood
window size in grid cells, and \(\mathcal{N}_w(i)\) is the \(w \times w\)
neighborhood centered at grid point \(i\). \(V_j\) is the valid-data indicator,
and \(\Omega_w\) denotes grid points with at least one valid point in the
neighborhood.

\begin{equation}
\begin{gathered}
\begin{aligned}
\rho_r
&=
\frac{
  \sum_{i \in \Omega}
  \left(
    \hat{A}_{r,i}
    -
    \overline{\hat{A}}_r
  \right)
  \left(
    A_i
    -
    \overline{A}
  \right)
}{
  \sqrt{
    \sum_{i \in \Omega}
    \left(
      \hat{A}_{r,i}
      -
      \overline{\hat{A}}_r
    \right)^2
  }
  \sqrt{
    \sum_{i \in \Omega}
    \left(
      A_i
      -
      \overline{A}
    \right)^2
  }
},
\end{aligned}
\\[0.5em]
\begin{aligned}
\mathrm{Bias}_r
&=
100
\times
\frac{
  \sum_{i \in \Omega}
  b_i \hat{A}_{r,i}
  -
  \sum_{i \in \Omega}
  b_i A_i
}{
  \sum_{i \in \Omega}
  b_i A_i
},
\qquad
b_i = \cos(\phi_i).
\end{aligned}
\end{gathered}
\end{equation}

where \(\rho_r\) is the spatial Pearson correlation between forecast member \(r\)
and IMERG, and \(\mathrm{Bias}_r\) is the cosine-latitude-weighted regional
precipitation bias in percent. The overbars denote unweighted spatial means over
valid grid points in the regional domain, and \(\phi_i\) is the latitude of grid
point \(i\).

\begin{equation}
\begin{gathered}
\begin{aligned}
p_i(q)
&=
\frac{1}{|\mathcal{M}|}
\sum_{r \in \mathcal{M}}
\mathbf{1}
\left[
  \hat{A}_{r,i} \ge q
\right],
\end{aligned}
\\[0.5em]
\begin{aligned}
\mathrm{BS}_{\mathrm{region}}(q)
&=
\frac{1}{N}
\sum_{i \in \Omega}
\left(
  p_i(q)
  -
  O_i(q)
\right)^2 .
\end{aligned}
\end{gathered}
\end{equation}

where \(p_i(q)\) is the ensemble exceedance probability of 72-h precipitation
exceeding threshold \(q\), and \(\mathrm{BS}_{\mathrm{region}}(q)\) is the
unweighted regional Brier Score over the event domain. In this case study,
\(q=50\) mm.

\section{Dataset}   
\subsection{ERA5 and IMERG}
\noindent ERA5 is the fifth-generation reanalysis produced by ECMWF, which provides a globally complete and physically consistent reconstruction of the atmosphere by assimilating a wide range of diverse observations with a state-of-the-art numerical weather prediction system \cite{ref-book9}. ERA5 offers hourly data at a horizontal resolution of 0.25° (~31 km) on a global (721 × 1440) latitude–longitude grid. Its extensive temporal coverage and high accuracy make it the most widely used benchmark dataset for evaluating weather and climate models.

In this study, IMERG precipitation (version 07) is adopted as the label dataset, which is available globally at 0.1° grid spacing (approximately 10 km) and 30 min temporal resolution \cite{ref-book11}. Three versions of IMERG are available with varying latency: IMERG Early, IMERG Late, and IMERG Final. IMERG Final includes the most available passive microwave (PMW) retrievals and bias correction and is therefore selected here for analysis. In addition, IMERG provides an observationally based precipitation estimate with finer sub-daily temporal sampling than ERA5, particularly for representing the diurnal cycle, whereas ERA5 tends to underestimate sub-daily precipitation variability \cite{ref-book11, ref-book12}. Since our experiments are based on 6-hourly accumulated precipitation (00, 06, 12, and 18 UTC), IMERG’s improved capability at sub-daily scales makes it a more suitable choice of label than ERA5 for global precipitation forecasting.

The CSU-PCAST model ingests a comprehensive set of atmospheric variables from ERA5 reanalysis, comprising 57 input channels in total. For the upper-air fields, six key variables are selected across eight pressure levels (200, 250, 300, 400, 500, 600, 700, and 850 hPa), including geopotential (Z), temperature (T), zonal wind (U), meridional wind (V), specific humidity (Q), and vertical velocity (W). These variables capture the thermodynamic and dynamic structures of the atmosphere, providing essential information on circulation, moisture transport, and vertical motion associated with precipitation processes. Several near-surface and surface variables are included to better constrain boundary-layer and column water conditions. These variables consist of 2-meter temperature (2T), 2-meter dewpoint temperature (2D), 10-meter winds (U10, V10), mean sea-level pressure (MSL), convective available potential energy (CAPE), total column water vapor (TCWV), surface pressure (SP), and top-layer soil moisture (SWVL1). Together, these variables provide critical information on near-surface thermodynamics, water vapor availability, and land–atmosphere coupling, all of which play vital roles in precipitation development and evolution.

For training, CSU-PCAST makes use of 21 years of reanalysis data covering 1998–2018. The year 2019 is held out for validation.
\subsection{GFS and GEFS}
Inference and testing are performed using GFS analysis, with GEFS serving as the operational baseline for comparison. The GFS and the GEFS are two key operational prediction systems developed at the National Centers for Environmental Prediction (NCEP).
GFS serves as the deterministic backbone, providing global forecasts of atmospheric and wave conditions at ~13 km horizontal resolution, with 127 vertical layers, run four times per day (00, 06, 12, and 18 UTC) and extending out to 16 days \cite{ref-gfs}. It uses the Finite Volume Cubed (FV3) dynamical core, coupled with the MOM6 ocean and CICE6 sea ice models, and is initialized through the hybrid ensemble, variational assimilation scheme of the Global Data Assimilation System (GDAS).

Building on GFS, the GEFS provides ensemble-based probabilistic forecasts. GEFS was first implemented in 1992 with a small number of perturbed members generated by the breeding vector method, and gradually increased its ensemble size and complexity over the years. By the mid-2000s, GEFS operated with 20 perturbed members plus one control, cycling every 6 hours and extending forecasts to 16 days \cite{ref-book-gefs}. A major upgrade in October 2020 (GEFSv12) expanded the ensemble to 30 perturbed members plus one control, with a forecast length of 35 days and horizontal resolutions of 0.25°, 0.5°, and 1.0°. Initial conditions are provided by the operational hybrid ensemble Kalman filter (EnKF) system, and stochastic physics schemes, including the stochastically perturbed parameterization tendency (SPPT) scheme and the stochastic kinetic energy backscatter (SKEB) scheme, are used to represent model uncertainties \cite{zhou2022gefs, palmer2009stochastic, berner2009spectral}.

For consistency with our experimental setup, the GEFS baseline used for comparison is taken at its native 0.5° resolution. Although a 0.25° GEFS atmospheric product is also available, it only extends to forecast hour 240, whereas the 0.5° product supports forecast integrations beyond day 10 and is therefore suitable for our 15-day evaluation. GEFS forecasts are used as raw model output without additional post-processing. The raw GEFS precipitation fields are bilinearly interpolated to the 0.25° latitude--longitude grid prior to verification against IMERG observations. To assess whether this interpolation affects the comparison, we evaluated the interpolated 0.5° GEFS forecasts against the native 0.25° GEFS product over their common forecast range in 2023 using CSI and FSS. The two versions produced nearly identical scores, suggesting that the interpolation-related error is negligible for the metrics considered in this study. For this study, we use 30 GEFS ensemble members to match the CSU-PCAST ensemble size, with forecast integrations extending to 15 days.

\begin{table}[t]
\centering
\begin{threeparttable}
\caption{Summary of input variables used by CSU\textendash PCAST.}
\label{tab:variables}
\renewcommand{\arraystretch}{1.1}
\setlength{\tabcolsep}{6pt}

\begin{tabular}{@{}lllc@{}}
\toprule
\textbf{Type} & \textbf{Variable name} & \textbf{Abbreviation} & \textbf{Role} \\
\midrule
\multirow[t]{6}{*}{Upper-air variables*}
 & Geopotential                     & Z     & Input and Predicted \\
 & Temperature                                  & T     & Input and Predicted \\
 & U component of wind                          & U     & Input and Predicted \\
 & V component of wind                          & V     & Input and Predicted \\
 & Specific humidity                            & Q     & Input and Predicted \\
 & Vertical velocity                            & W     & Input and Predicted \\
\midrule
\multirow[t]{9}{*}{Surface variables}
 & 2-meter temperature                          & 2T    & Input and Predicted \\
 & 2-meter dewpoint temperature                 & 2D    & Input and Predicted \\
 & 10-meter u wind component                    & U10   & Input and Predicted \\
 & 10-meter v wind component                    & V10   & Input and Predicted \\
 & Mean sea-level pressure                      & MSL   & Input and Predicted \\
 & Convective available potential energy        & CAPE  & Input and Predicted \\
 & Total column water vapor                     & TCWV  & Input and Predicted \\
 & Surface pressure                             & SP    & Input and Predicted \\
 & Top-layer soil moisture                      & SWVL1 & Input and Predicted \\
 & Total precipitation                          & TP    & Predicted (6h) \\
\midrule
\multirow[t]{3}{*}{Geographical}
 & Land–sea mask                                & LSM   & Input \\
 & Soil type                                    & SOIL  & Input \\
 & Topography (orography)                       & ORO   & Input \\
\bottomrule
\end{tabular}

\begin{tablenotes}[flushleft]
\footnotesize
\item[*] Upper-air fields are taken at 8 pressure levels: 200, 250, 300, 400, 500, 600, 700, and 850\,hPa.
\end{tablenotes}
\end{threeparttable}
\end{table}

\section{Results}
\noindent This section evaluates the proposed ensemble forecasting framework against the operational GEFS ensemble over the full year of 2023. Forecasts are generated from initial conditions at 00, 06, 12, and 18 UTC, with CSU-PCAST initialized from the corresponding operational GFS analysis at each cycle. CSU-PCAST produces 30 ensemble members, matching the operational GEFS configuration. For each initial time, CSU-PCAST requires approximately 15 minutes on three NVIDIA H100 GPUs to generate a full 15-day forecast, corresponding to 60 forecast steps at 6-h intervals, with each GPU producing 10 ensemble members. By comparison, operational GEFS production involves a substantially larger physics-based ensemble workflow and typically completes on the order of a few hours per forecast cycle, approximately 3 h in NCEP production monitoring \cite{ncep_prodstat}. The results are presented in three parts: deterministic verification, probabilistic verification, and a case study of the Sanba extreme precipitation event over South China in October 2023.

\subsection{Deterministic Verification}
\begin{figure*}[htbp]
\centerline{\includegraphics[width=1\textwidth]{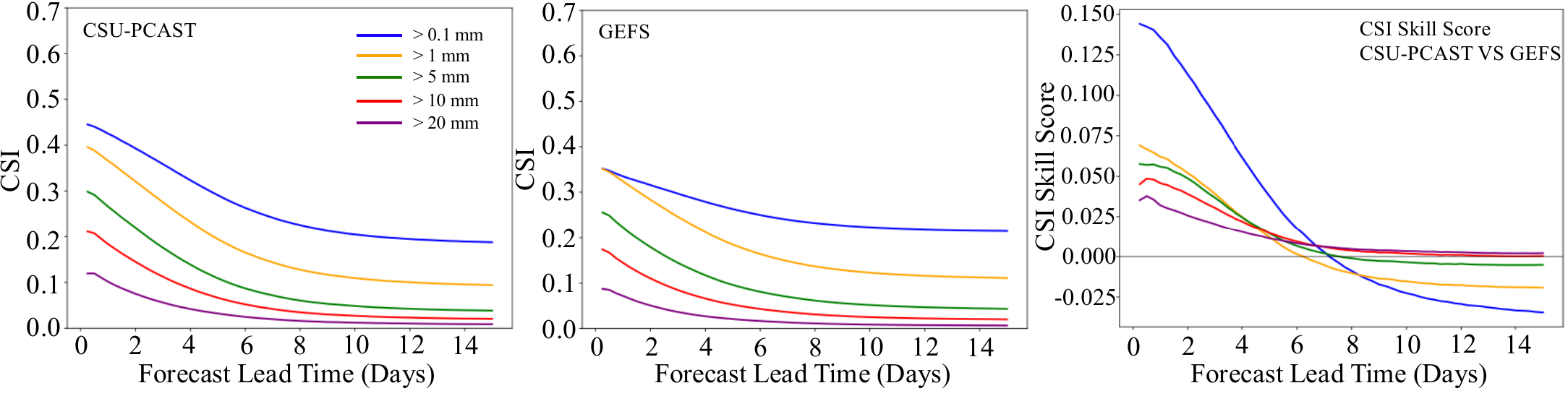}}
    \caption{Member-level categorical verification of precipitation forecasts over the full year of 2023. CSI for CSU-PCAST (left) and GEFS (middle), and CSISS of CSU-PCAST relative to GEFS (right). CSI is evaluated across precipitation thresholds and forecast lead times based on the individual ensemble members. Positive CSISS values indicate that CSU-PCAST outperforms GEFS.}
    \label{fig:csi_metrics}
\end{figure*}

\noindent This subsection compares the deterministic precipitation forecast performance of CSU-PCAST and GEFS. Fig.~\ref{fig:csi_metrics} shows the member-level CSI and CSISS over the 2023 evaluation period. For both CSU-PCAST and GEFS, CSI decreases with forecast lead time, with higher scores at lighter precipitation thresholds and lower scores for heavier precipitation events. At short lead times, CSU-PCAST achieves higher CSI than GEFS across the evaluated thresholds, especially for light precipitation, as indicated by the positive CSISS. This advantage weakens as lead time increases. For most thresholds, CSISS approaches zero by forecast days 5--7, and it becomes slightly negative at longer lead times for the lightest thresholds. These results indicate that CSU-PCAST provides a clearer deterministic categorical forecast advantage at shorter lead times, while the advantage is limited at longer leads.

\begin{figure}[!htbp]
\centering
\includegraphics[width=1\textwidth]{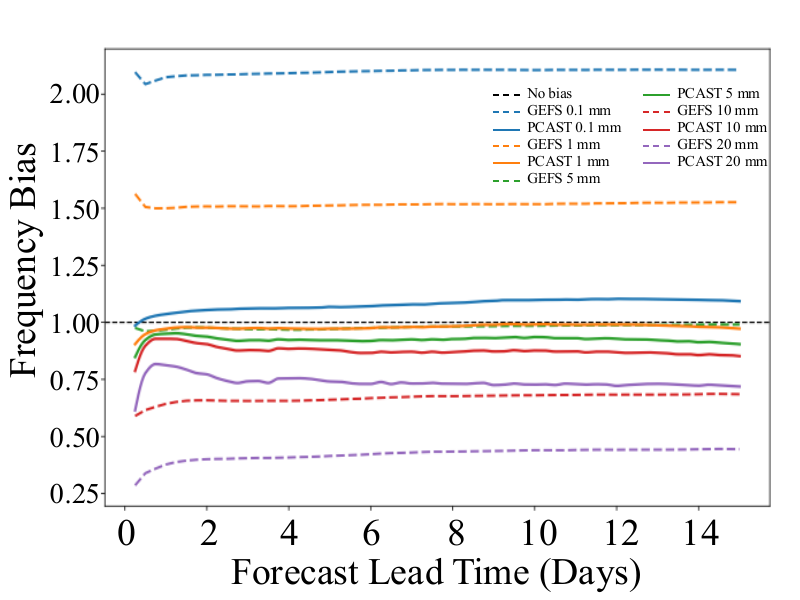}
\caption{Ensemble-pooled precipitation frequency bias of CSU-PCAST and GEFS relative to IMERG for forecasts over the full year of 2023. Bias is computed as the ratio of pooled forecast event counts to pooled observed IMERG event counts across all valid initializations, ensemble members, and grid points. Solid lines denote CSU-PCAST and dashed lines denote GEFS; colors indicate precipitation thresholds. A bias of 1 indicates no frequency bias.}
\label{fig:bias}
\end{figure}

Fig.~\ref{fig:bias} shows the ensemble-pooled precipitation frequency bias verified against IMERG. Values greater than one indicate overforecasting, while values less than one indicate underforecasting. GEFS exhibits a strong wet bias for light precipitation, with bias values near 2.0 at the 0.1 mm threshold and near 1.5 at the 1 mm threshold throughout the forecast range. Its bias is close to one for moderate precipitation near the 5 mm threshold, but GEFS substantially underforecasts events at the 10 and 20 mm thresholds. This dry bias is strongest at the earliest lead times and becomes slightly less pronounced with increasing lead time, although it remains below one throughout the forecast range. Compared with GEFS, CSU-PCAST has frequency bias values closer to one for light and moderate precipitation thresholds. At the lightest thresholds, CSU-PCAST reduces the strong GEFS wet bias and remains near unity, although it becomes slightly wet-biased at longer leads. At the 10 and 20 mm thresholds, both systems underforecast event frequency, but CSU-PCAST maintains higher bias values than GEFS, indicating a weaker dry bias. Overall, the ensemble-pooled frequency-bias evaluation suggests that CSU-PCAST improves the calibration of precipitation occurrence frequency relative to GEFS for the thresholds considered here, especially by reducing the excessive frequency of light precipitation and the underforecasting at the 10 and 20 mm thresholds.

\begin{figure*}[htbp]
\centerline{\includegraphics[width=1\textwidth]{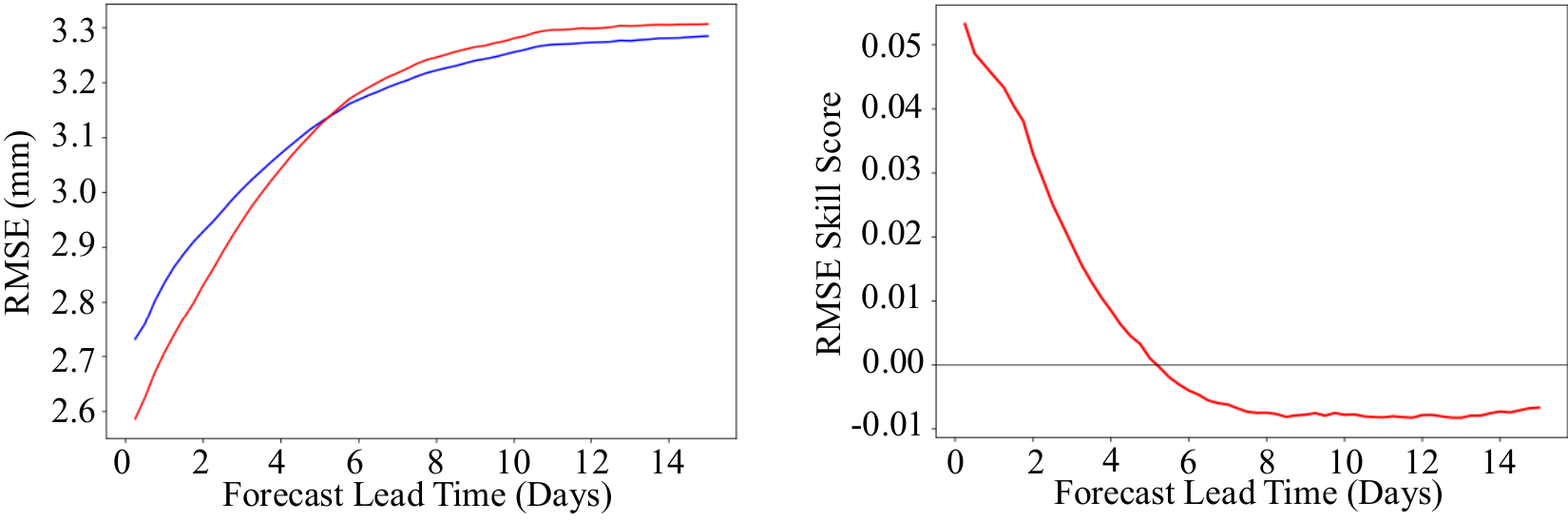}}
\centerline{\includegraphics[width=0.25\textwidth]{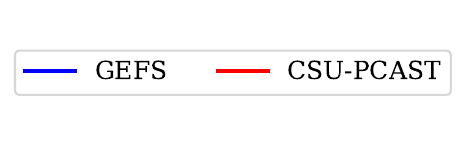}}
    \caption{Forecast-lead dependence of deterministic RMSE during the 2023 evaluation period. RMSE for CSU-PCAST and GEFS (left). RMSE skill score of CSU-PCAST relative to GEFS, with positive values indicating improvement over GEFS (right).}
    \label{fig:RMSE}
\end{figure*}

Fig.~\ref{fig:RMSE} shows the forecast-lead dependence of deterministic RMSE during the 2023 evaluation period. For both CSU-PCAST and GEFS, RMSE increases with lead time, with the fastest growth occurring during the first several forecast days. CSU-PCAST has lower RMSE than GEFS at the earliest lead times, and the corresponding RMSE skill score remains positive through roughly the first five forecast days. The largest improvement occurs at the shortest leads, after which the skill score steadily decreases toward zero. After about forecast days 5--6, the RMSE skill score becomes slightly negative, indicating that the CSU-PCAST advantage in deterministic precipitation magnitude does not persist at longer lead times. Overall, CSU-PCAST improves short-range deterministic precipitation forecasts relative to GEFS, while its RMSE-based benefit is limited at longer leads.

\FloatBarrier

\subsection{Probabilistic Verification}
\begin{figure*}[htbp]
\centerline{\includegraphics[width=1\textwidth]{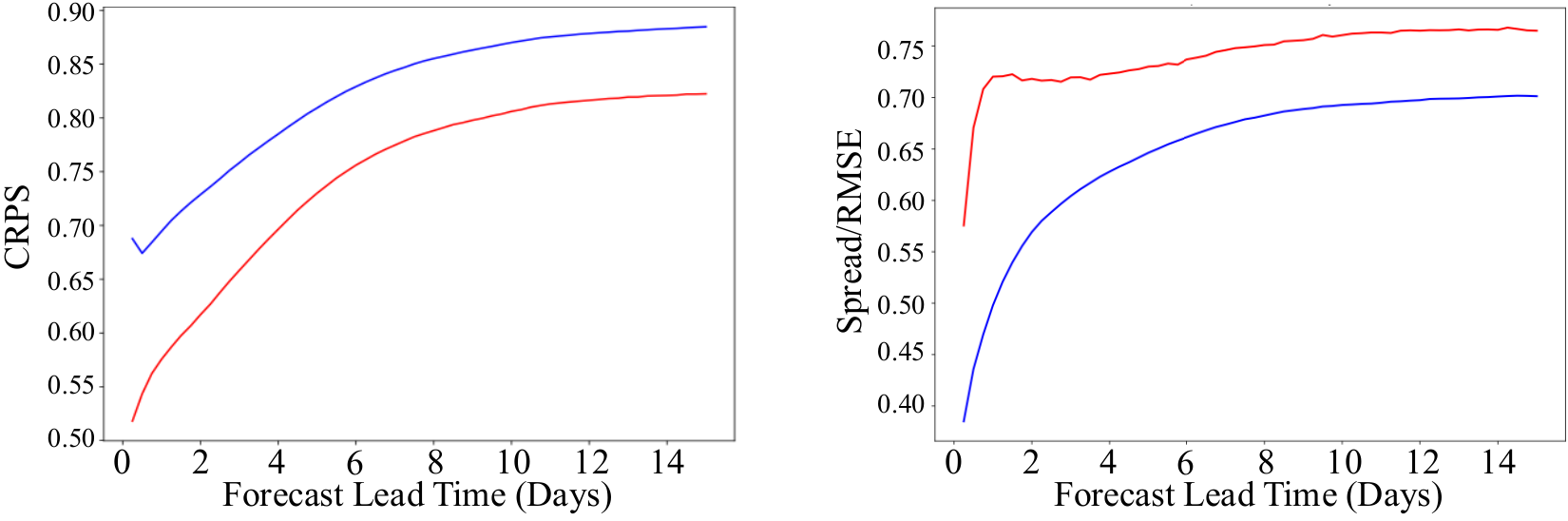}}
\centerline{\includegraphics[width=0.25\textwidth]{IMG/legend.pdf}}
    \caption{Forecast-lead dependence of probabilistic precipitation performance during the 2023 evaluation period. Latitude-weighted CRPS for CSU-PCAST and GEFS verified against IMERG, with lower values indicating better performance (left). Spread-to-RMSE ratio for CSU-PCAST and GEFS, with values closer to one indicating better agreement between ensemble spread and forecast error (right).}
    \label{fig:crps_ratio}
\end{figure*}

\noindent This subsection compares the probabilistic precipitation forecast performance of CSU-PCAST and GEFS. Fig.~\ref{fig:crps_ratio} shows the forecast-lead dependence of CRPS and the spread-to-RMSE ratio during the 2023 evaluation period. For both CSU-PCAST and GEFS, CRPS increases with forecast lead time, indicating a gradual degradation in probabilistic precipitation forecast quality. CSU-PCAST has lower CRPS than GEFS throughout the forecast range, with the largest difference at the earliest lead times. The gap decreases with increasing lead time but remains evident at longer leads. The spread-to-RMSE ratio remains below one for both systems, indicating that both ensembles are underdispersive relative to their forecast errors. CSU-PCAST has larger spread-to-RMSE ratios than GEFS across all lead times, suggesting better ensemble dispersion. The CSU-PCAST ratio increases rapidly during the first forecast day and then remains near 0.7--0.8, whereas the GEFS ratio increases more gradually and stays lower throughout the forecast range. These results indicate that CSU-PCAST provides improved probabilistic precipitation forecasts relative to GEFS in terms of both CRPS and ensemble dispersion, although both systems remain underdispersive.

\begin{figure}[htbp]
\centerline{\includegraphics[width=0.7\textwidth]{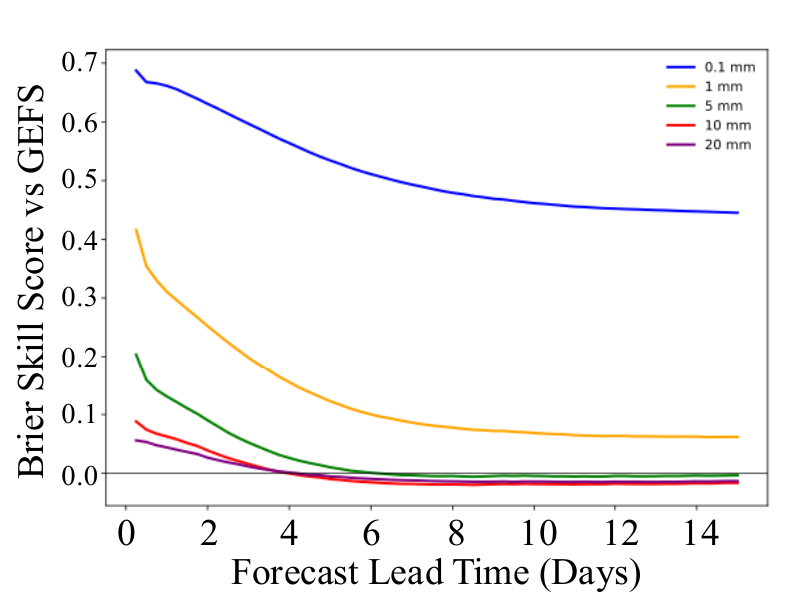}}
    \caption{Forecast-lead dependence of BSS for CSU-PCAST relative to GEFS during the 2023 evaluation period. BSS is shown for precipitation thresholds of 0.1, 1, 5, 10, and 20 mm, with positive values indicating improvement over GEFS.}
    \label{fig:BSS}
\end{figure}

Fig.~\ref{fig:BSS} shows the Brier Skill Score of CSU-PCAST relative to GEFS for different precipitation thresholds during the 2023 evaluation period. CSU-PCAST shows positive BSS at short lead times for all evaluated thresholds, indicating improved probabilistic event forecasts relative to GEFS. The largest skill occurs for light precipitation, especially at the 0.1 mm threshold, where BSS remains strongly positive throughout the forecast range. Skill decreases as the precipitation threshold increases. For the 1 mm threshold, BSS remains positive but declines steadily with lead time, while for the 5 mm threshold it approaches zero after about forecast days 5--6. At the 10 and 20 mm thresholds, BSS becomes slightly negative at longer lead times, indicating limited or no probabilistic skill improvement over GEFS for heavier precipitation events. Overall, CSU-PCAST provides the clearest BSS improvement for light precipitation and short lead times, with weaker gains for heavier thresholds and longer forecasts.

\begin{figure}[!htbp]
    \centering
    \includegraphics[width=0.75\textwidth,trim=0 10 0 10,clip]{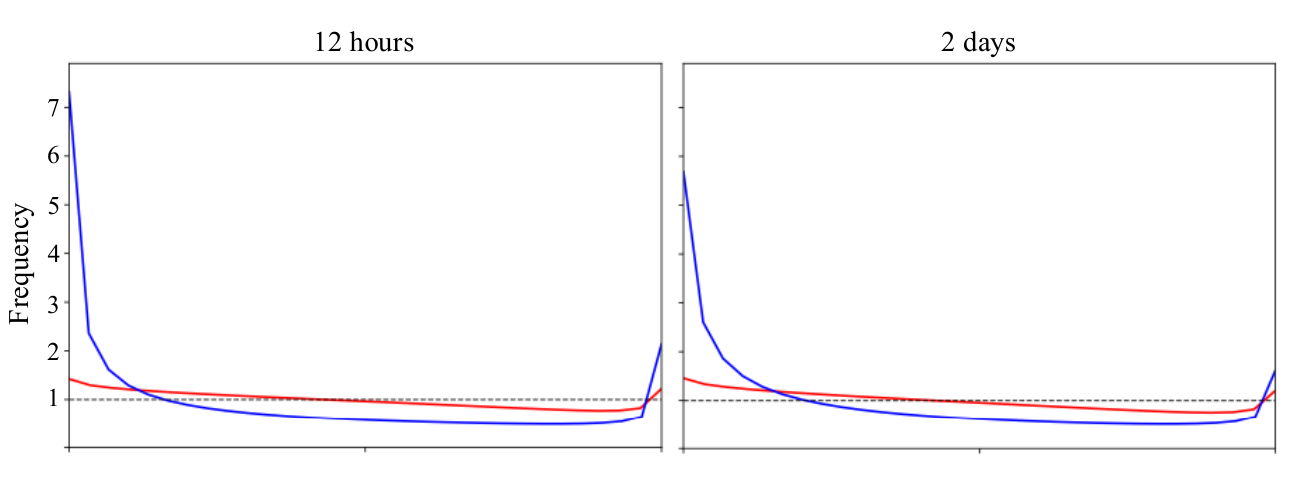}\\[-0.4em]
    \includegraphics[width=0.75\textwidth,trim=0 0 0 0,clip]{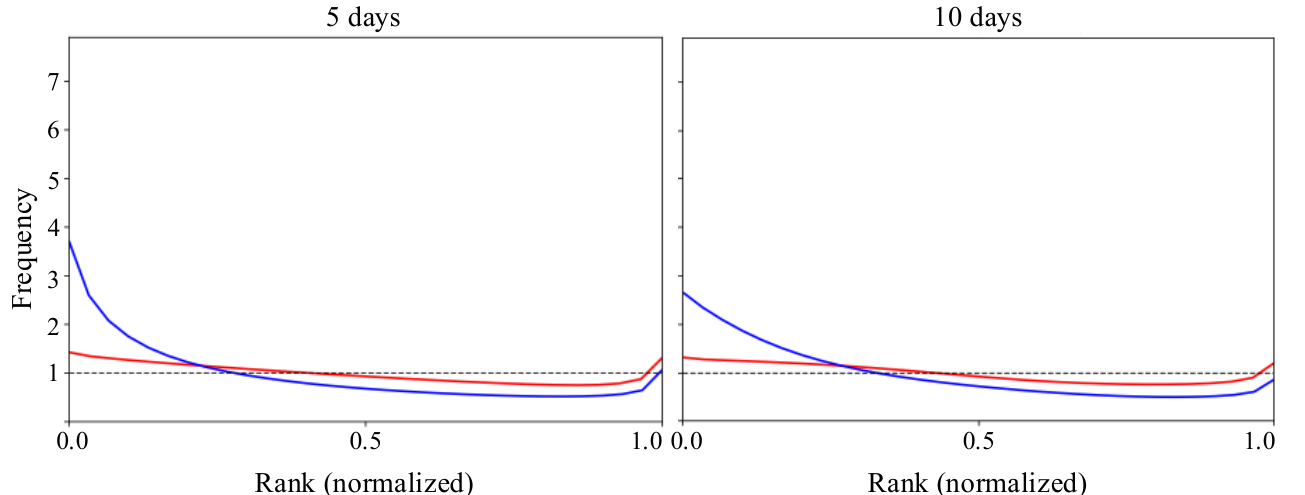}\\[-0.4em]
    \includegraphics[width=0.20\textwidth]{IMG/legend.pdf}
    \caption{PIT-style rank histograms for CSU-PCAST and GEFS precipitation forecasts at selected lead times during the 2023 evaluation period. The rank is normalized by ensemble size, with values near 0 indicating observations below the ensemble and values near 1 indicating observations above the ensemble. Frequencies are normalized by the expected count under a uniform rank distribution, and the dashed line marks the ideal value of one.}
    \label{fig:rank_hist}
\end{figure}

Fig.~\ref{fig:rank_hist} shows probability integral transform (PIT) style rank histograms for CSU-PCAST and GEFS at selected forecast lead times. A reliable ensemble would produce a nearly uniform rank distribution, corresponding to normalized frequencies close to one. GEFS shows a pronounced concentration near the lowest ranks, especially at short lead times, indicating that observations often fall near or below the lower end of the ensemble distribution. This pattern is consistent with a tendency toward high precipitation forecasts relative to IMERG. The GEFS rank distribution becomes less sharply peaked with increasing lead time, but it remains noticeably nonuniform across the forecast range. In contrast, CSU-PCAST produces rank histograms that are closer to the ideal uniform distribution, with weaker endpoint concentrations. Some departures from uniformity remain, including modest excess frequencies at low ranks and slight edge effects, but the overall rank structure suggests better ensemble reliability than GEFS.

\begin{figure}[!htbp]
\centering
\includegraphics[width=0.65\textwidth]{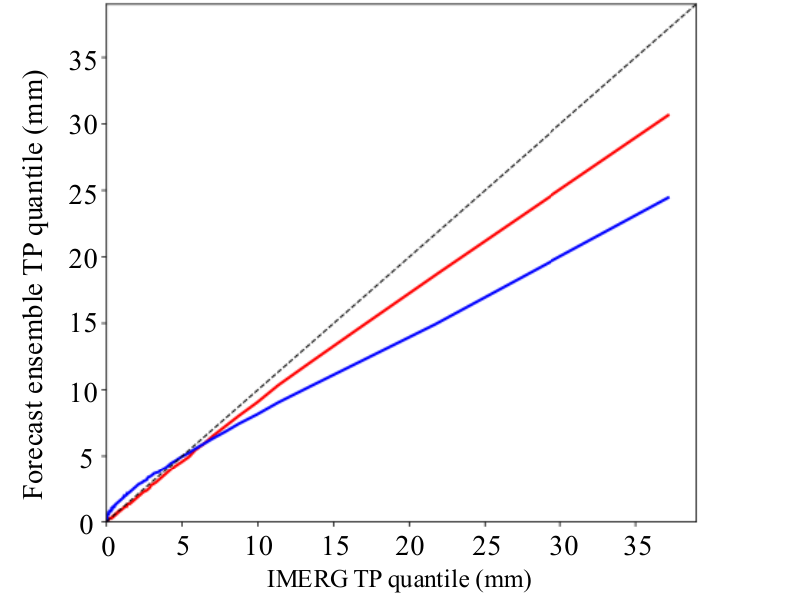}\\[-0.3em]
\includegraphics[width=0.25\textwidth]{IMG/legend.pdf}
\caption{Q-Q plot of ensemble precipitation forecasts against IMERG during the 2023 evaluation period. Forecast precipitation quantiles from CSU-PCAST and GEFS are plotted against the corresponding IMERG precipitation quantiles, with the dashed black line indicating the one-to-one reference.}
\label{fig:QQ}
\end{figure}

Fig.~\ref{fig:QQ} compares the precipitation quantiles of CSU-PCAST and GEFS against the corresponding IMERG quantiles. Both systems are close to the one-to-one line at lower precipitation quantiles, indicating relatively consistent behavior for light precipitation amounts. At higher quantiles, both systems fall below the reference line, showing that the upper tail of the precipitation distribution is underestimated relative to IMERG. CSU-PCAST remains closer to the one-to-one line than GEFS across most of the distribution, especially at moderate to high quantiles. This indicates that CSU-PCAST better represents the precipitation intensity distribution, although it still underestimates the highest precipitation quantiles.

\FloatBarrier

\subsection{Case Study: The Sanba Extreme Precipitation Event}
\noindent The Sanba extreme precipitation event is used to evaluate the performance of CSU-PCAST and GEFS for a high-impact regional rainfall case. Verification is conducted over South China within \(17^\circ\)--\(25^\circ\)N and \(104^\circ\)--\(114^\circ\)E, using 72-h accumulated precipitation from 2023-10-17 06 UTC to 2023-10-20 06 UTC. Forecasts are initialized at 2023-10-14 06 UTC, corresponding to lead days 3--5. The event threshold is defined as 50 mm per 72 h.

\begin{figure}[!htbp]
\centering
\includegraphics[width=0.95\linewidth]{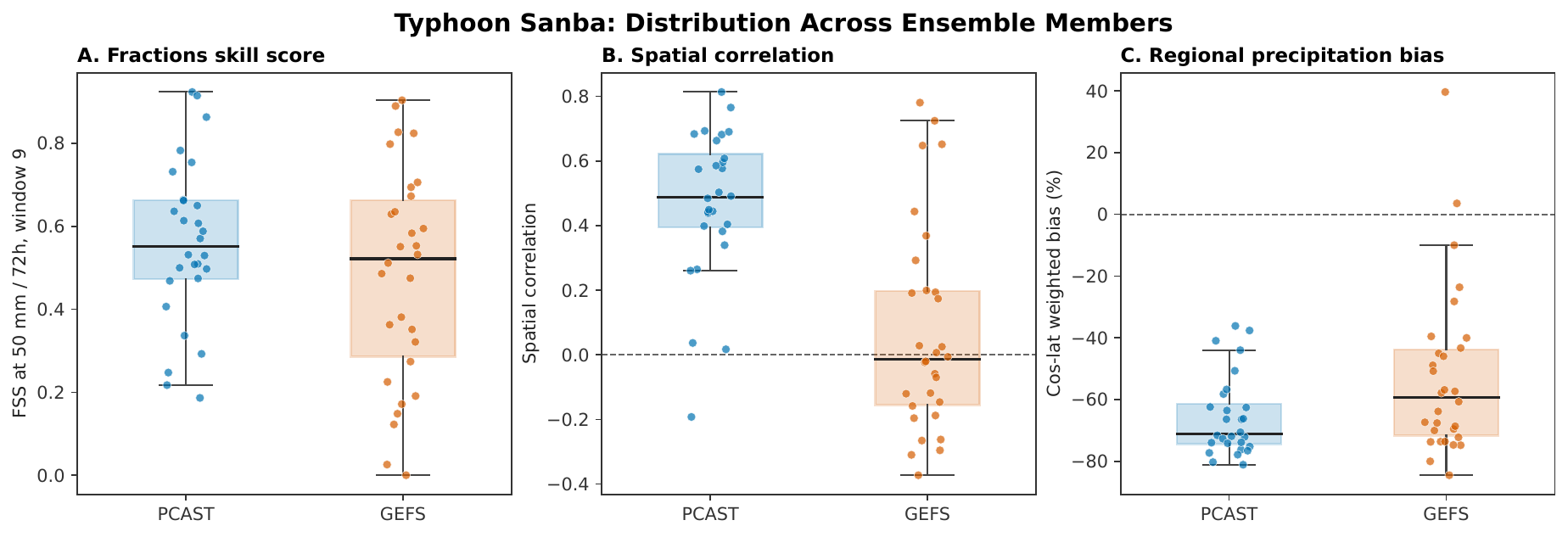}

\vspace{0.5em}

\includegraphics[width=0.95\linewidth]{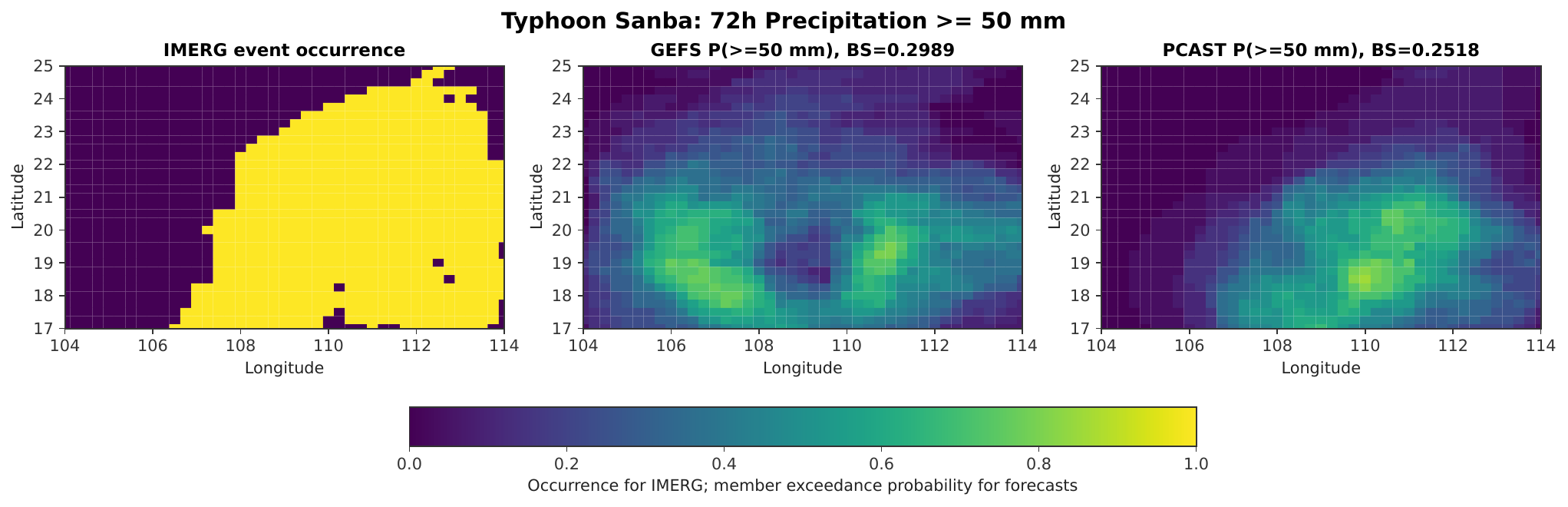}

\caption{Event-based verification of 72-h precipitation forecasts for Typhoon Sanba. The upper panel shows member-level distributions of FSS at the 50 mm per 72 h threshold, spatial correlation with IMERG, and cosine-latitude-weighted regional precipitation bias. The lower panel shows observed occurrence and ensemble probabilities of 72-h precipitation exceeding 50 mm; CSU-PCAST has a lower Brier score than GEFS (0.2518 versus 0.2989).}
\label{fig:sanba_event}
\end{figure}

The upper panel of Fig.~\ref{fig:sanba_event} summarizes the member-level verification metrics for the 72-h accumulated precipitation forecasts. CSU-PCAST has a higher mean FSS than GEFS at the 50 mm threshold, with mean values of 0.560 and 0.481, respectively. The median FSS values are more similar, but CSU-PCAST has a higher lower quartile, indicating fewer low-skill members. CSU-PCAST also shows a substantially higher spatial correlation with IMERG, with a mean correlation of 0.468 compared with 0.071 for GEFS. This indicates that CSU-PCAST better captures the spatial structure of the Sanba precipitation field in this case. However, both systems underestimate regional precipitation totals on average. The mean cosine-latitude-weighted precipitation bias is \(-65.6\%\) for CSU-PCAST and \(-52.6\%\) for GEFS, indicating that CSU-PCAST has the larger dry bias in total precipitation despite its stronger spatial agreement.

The lower panel of Fig.~\ref{fig:sanba_event} compares the ensemble exceedance probability fields for the 50 mm per 72 h threshold. IMERG exceeds this threshold over 713 of 1353 grid cells, corresponding to 52.7\% of the verification region. CSU-PCAST has a lower Brier score than GEFS, with values of 0.252 and 0.299, respectively, indicating better probabilistic agreement with the observed event mask. GEFS assigns nonzero exceedance probability over a larger fraction of the domain and has a higher domain-mean probability, but this broader probability coverage does not translate into a lower Brier score. Overall, the Sanba case shows that CSU-PCAST provides better spatial structure and probabilistic event guidance than GEFS for this lead period, while both systems remain too dry relative to IMERG and CSU-PCAST underestimates the regional precipitation total more strongly.

\FloatBarrier

\section{Discussion}
\noindent The annual verification results show that CSU-PCAST provides a clear improvement over GEFS for several aspects of precipitation forecasting, but the improvement is not uniform across lead time, threshold, and metric. In the deterministic verification, CSU-PCAST performs best at short lead times. The CSI and RMSE skill scores are positive during the first several forecast days, indicating better categorical precipitation occurrence and lower precipitation-magnitude errors than GEFS. These advantages weaken with increasing lead time and become small or slightly negative after about forecast days 5--7, suggesting that the deterministic benefit of CSU-PCAST is mainly concentrated in the short range.

The frequency-bias results provide a more nuanced view of these deterministic improvements. GEFS strongly overforecasts light precipitation while underforecasting events at the 10 and 20 mm thresholds. CSU-PCAST reduces both the excessive frequency of light precipitation and the dry bias at these heavier annual thresholds. This behavior suggests that CSU-PCAST produces a more balanced precipitation occurrence distribution for the thresholds evaluated in the annual verification. At the same time, the improvement should not be interpreted as a complete correction of precipitation intensity, because the Q-Q analysis shows that both systems still underestimate the upper tail of the precipitation distribution relative to IMERG.

The probabilistic verification further indicates that CSU-PCAST gives a better representation of ensemble precipitation uncertainty than GEFS. CSU-PCAST has lower CRPS across the forecast range and higher BSS for most thresholds at short lead times, with the clearest gains for light precipitation. The spread-to-RMSE ratio and PIT-style rank histograms also show improved ensemble dispersion and reliability relative to GEFS. However, both systems remain underdispersive, and the BSS advantage decreases for heavier precipitation thresholds and longer lead times. These results indicate that CSU-PCAST improves probabilistic forecast quality, but the ensemble spread is still too narrow relative to the forecast error. The pattern of improvement suggests that CSU-PCAST is particularly effective at correcting broad precipitation occurrence and reducing some of the systematic light-precipitation bias present in GEFS. However, the weaker gains at longer lead times and higher thresholds indicate that the model still has difficulty maintaining accurate precipitation intensity and ensemble spread as forecast uncertainty grows. This distinction is important because precipitation occurrence, spatial placement, intensity, and uncertainty calibration can improve at different rates. Additional latitude-band diagnostics further show that these improvements are not spatially uniform (Appendix~\ref{sec:latband}). CSU-PCAST maintains lower CRPS than GEFS in both the tropics and midlatitudes, but the categorical skill scores show stronger dependence on latitude band, threshold, and lead time. In particular, BSS improvements are most persistent for light precipitation, while member-level CSI skill becomes more mixed at longer leads and higher thresholds. These results indicate that the global aggregate scores summarize the broad advantage of CSU-PCAST, but they do not fully capture regional-regime differences in precipitation skill.

The Sanba case study provides an event-scale perspective that complements the annual verification. For the 72-h accumulated precipitation associated with Sanba, CSU-PCAST achieves higher FSS and much higher spatial correlation than GEFS, indicating a better representation of the observed precipitation structure. It also produces a lower Brier score for the 50 mm exceedance probability field. However, both systems underestimate the regional precipitation total, and CSU-PCAST has a larger negative regional bias than GEFS in this case. This contrast shows that better spatial placement and probabilistic event guidance do not necessarily imply better regional precipitation volume. Together, the annual and case-study results suggest that CSU-PCAST improves several important aspects of precipitation forecast skill relative to GEFS, while high-end precipitation intensity and ensemble dispersion remain areas for further improvement.

\section{Limitations}
\noindent Several limitations should be considered when interpreting these results. Although CSU-PCAST shows improved annual performance over GEFS for several global precipitation metrics, this advantage is not uniform across regions, verification datasets, or spatial scales. As an additional regional check, we evaluated ensemble-mean precipitation over the 	contiguous United States (CONUS) domain (\(20^\circ\)--\(58^\circ\)N, \(226^\circ\)--\(300^\circ\)E) against Stage IV precipitation. The CSI and FSS differences between CSU-PCAST and GEFS show mixed behavior, with positive values indicating better CSU-PCAST performance. In this CONUS verification, CSU-PCAST does not consistently outperform GEFS across thresholds and lead times. This suggests that the relative skill of CSU-PCAST can depend on the verification region and reference dataset.

The spatial-activity diagnostics also indicate that CSU-PCAST produces smoother precipitation fields than both IMERG and GEFS (Appendix~\ref{sec:spatial_activity_appendix}). In the global 6-h precipitation evaluation, both forecast systems have weaker precipitation gradients than IMERG, but the reduction is larger for CSU-PCAST. Across lead times, CSU-PCAST retains about 39\% of the IMERG mean gradient magnitude, compared with about 60\% for GEFS. Similar behavior is found for the upper-tail gradient magnitude, where CSU-PCAST retains about 44\% of the IMERG value and GEFS about 65\%. The high-gradient fraction is also lower in CSU-PCAST, indicating fewer sharp precipitation features. These results suggest that CSU-PCAST tends to damp small-scale precipitation structures, which may contribute to the underestimation of high-end precipitation quantiles and regional precipitation totals in some extreme events. This behavior may partly reflect the regression-based nature of the model, since optimization toward average error can favor smoother fields and suppress small-scale, high-amplitude precipitation features.

The evaluation period and reference datasets further limit the generality of the conclusions. The main verification is based on one year of forecasts and uses IMERG as the primary precipitation reference. Although IMERG provides consistent global coverage, it has retrieval uncertainty, particularly for intense and localized precipitation. The Stage IV analysis provides an independent regional check over CONUS, but it does not fully address robustness across other regions or years.

\section{Conclusion and Future Work}

\noindent This study evaluated CSU-PCAST as an ensemble precipitation forecasting framework against the operational GEFS ensemble over the full year of 2023. CSU-PCAST was initialized from operational GFS analyses and produced 30 ensemble members, matching the GEFS ensemble size. The annual verification showed that CSU-PCAST improves several aspects of precipitation forecast performance relative to GEFS, especially at short lead times. In deterministic verification, CSU-PCAST achieved higher CSI and lower RMSE during the first several forecast days, although these advantages weakened at longer leads. The frequency-bias analysis showed that CSU-PCAST reduced the strong GEFS wet bias for light precipitation and the dry bias at the 10 and 20 mm thresholds.

Probabilistic verification also showed consistent benefits from CSU-PCAST. The model achieved lower CRPS than GEFS across the forecast range and higher BSS for several thresholds, with the strongest gains for light precipitation and shorter lead times. Rank histograms and spread-to-RMSE ratios indicated improved ensemble reliability and dispersion relative to GEFS, although both systems remained underdispersive. The Q-Q analysis further showed that CSU-PCAST better represented the precipitation intensity distribution than GEFS across much of the range, while both systems underestimated the upper tail.

The Sanba extreme precipitation case demonstrated that CSU-PCAST can provide improved spatial structure and probabilistic event guidance for a high-impact regional event. CSU-PCAST achieved higher FSS, higher spatial correlation, and a lower Brier score than GEFS for the 50 mm per 72 h exceedance event. However, both systems underestimated regional precipitation totals, and CSU-PCAST showed a larger dry bias in this case. Additional diagnostics also indicated that CSU-PCAST produces smoother precipitation fields than IMERG and GEFS, suggesting that high-end precipitation intensity and small-scale precipitation structure remain important challenges.

Future work will focus first on improving precipitation forecasts over the CONUS region, where additional verification against Stage IV indicates that CSU-PCAST does not consistently outperform GEFS across thresholds and lead times. We will also explore bias-correction and calibration techniques to improve precipitation frequency, intensity, and ensemble reliability. These approaches may help reduce regional biases, improve the representation of high-end precipitation, and better calibrate ensemble probabilities while preserving the short-range forecast skill demonstrated in the annual evaluation.

\section*{Data availability}
\noindent GEFS forecast data were accessed and downloaded using the Herbie Python package, following the official documentation at \url{https://herbie.readthedocs.io/en/2024.5.0/user_guide/tutorial/model_notebooks/gefs.html}. GFS forecast data were retrieved via the Herbie Python package as described in the documentation at \url{https://herbie.readthedocs.io/en/2024.5.0/user_guide/tutorial/model_notebooks/gfs.html}. The ERA5 reanalysis dataset (0.25° grid) was obtained from the NSF NCAR GDEX repository (\url{https://gdex.ucar.edu/datasets/d633000/}), which contains modified Copernicus Climate Change Service information. IMERG precipitation products are publicly available from the NASA GES DISC archive at \url{https://gpm1.gesdisc.eosdis.nasa.gov/data/GPM_L3/GPM_3IMERGHH.07/}.

\section*{Code availability}
\noindent Pretrained model checkpoints and an inference pipeline for generating forecasts will be made available at \url{https://precipitation.engr.colostate.edu/}. The repository includes scripts for data preprocessing, model inference, and post-processing required to reproduce the forecasting results presented in this study. Training scripts are available from the corresponding author upon reasonable request.

\section*{Acknowledgments}
\noindent This research was supported by the National Oceanic and Atmospheric Administration (NOAA) through Grant NA25OARX405C0040-T1-01.
The authors thank Sergey Frolov at the NOAA Physical Sciences Laboratory for his feedback on this work. We also thank ECMWF for providing invaluable datasets to the research community.

\section*{Author contributions}
\noindent H.C. and J.T. conceived the essential ideas. T.X. performed the detailed analysis, including model training and evaluation, under the supervision of H.C.. K.M., T.S., and J.B. provided comments on the data analysis, results, and interpretation, and performed a comprehensive review to improve the depth and quality of this article. T.X. wrote the original manuscript. All authors contributed to the revisions of the paper.

\section*{Competing interests}
\noindent
The authors declare no competing interests.

\bibliographystyle{unsrt}  
\bibliography{references}  

\clearpage
\appendix 
\section{Appendix}

\renewcommand{\thefigure}{A\arabic{figure}}
\renewcommand{\thetable}{A\arabic{table}}
\renewcommand{\thesubsection}{A.\arabic{subsection}} 

\setcounter{figure}{0}
\setcounter{table}{0}
\setcounter{subsection}{0}

\FloatBarrier

\subsection{Sensitivity to Precipitation Training Target}
\label{sec:era5_appendix}

\begin{figure*}[htbp]
\centerline{\includegraphics[width=1\textwidth]{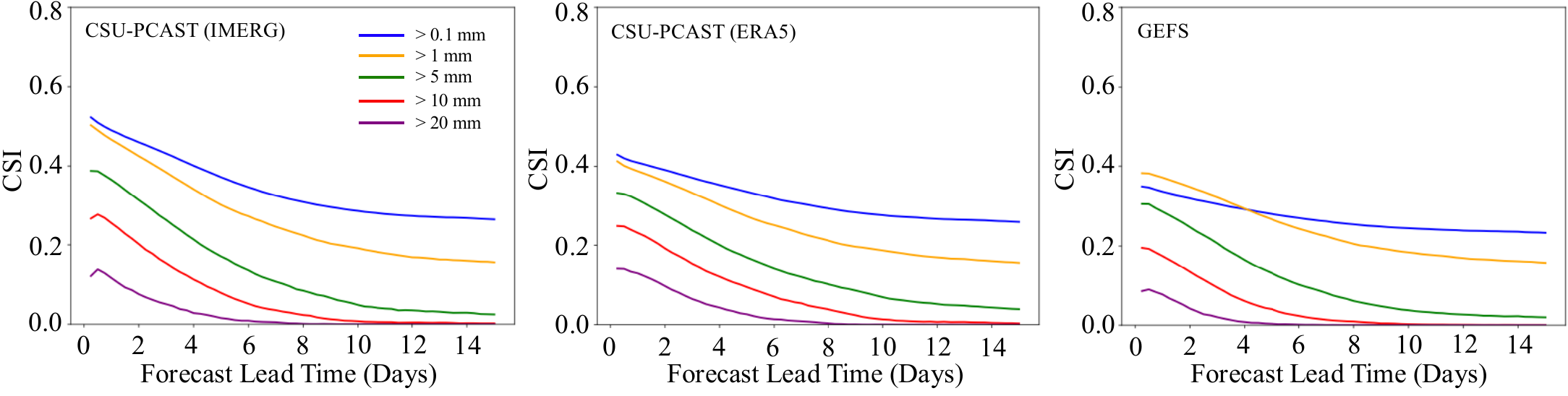}}
    \caption{Sensitivity of ensemble-mean CSI to the precipitation training target during the 2023 evaluation period. CSU-PCAST forecasts are initialized from operational GFS analyses and verified against IMERG for CSU-PCAST variants trained with IMERG precipitation labels (left) and ERA5 precipitation labels (middle). GEFS verified against IMERG is shown for reference (right). CSI is evaluated at precipitation thresholds of 0.1, 1, 5, 10, and 20 mm as a function of forecast lead time.}
    \label{fig:csi_internal_comp}
\end{figure*}

\noindent Fig.~\ref{fig:csi_internal_comp} compares ensemble-mean CSI for CSU-PCAST models trained with different precipitation targets. We use ensemble-mean CSI here as a compact diagnostic of the bulk categorical precipitation behavior associated with each training target.

When both models are initialized from operational GFS analyses and verified against IMERG, the CSU-PCAST model trained with IMERG precipitation labels achieves higher ensemble-mean CSI than the model trained with ERA5 precipitation labels across the evaluated thresholds. This indicates that using IMERG as the precipitation target better aligns the model output with the IMERG reference, including its precipitation occurrence, intensity distribution, and spatial structure. This result should be interpreted as an IMERG-consistency diagnostic, since IMERG is used both as the training target and as the verification reference.

The ERA5-trained CSU-PCAST still performs comparably to, and in many cases slightly better than, GEFS when all forecasts are verified against IMERG. This suggests that the CSU-PCAST architecture and training framework provide precipitation forecast benefits beyond the choice of IMERG as the precipitation label alone. At the same time, the difference between the IMERG-trained and ERA5-trained results highlights the sensitivity of precipitation skill estimates to the selected training target and verification dataset. 

\begin{figure*}[htbp]
\centerline{\includegraphics[width=1\textwidth]{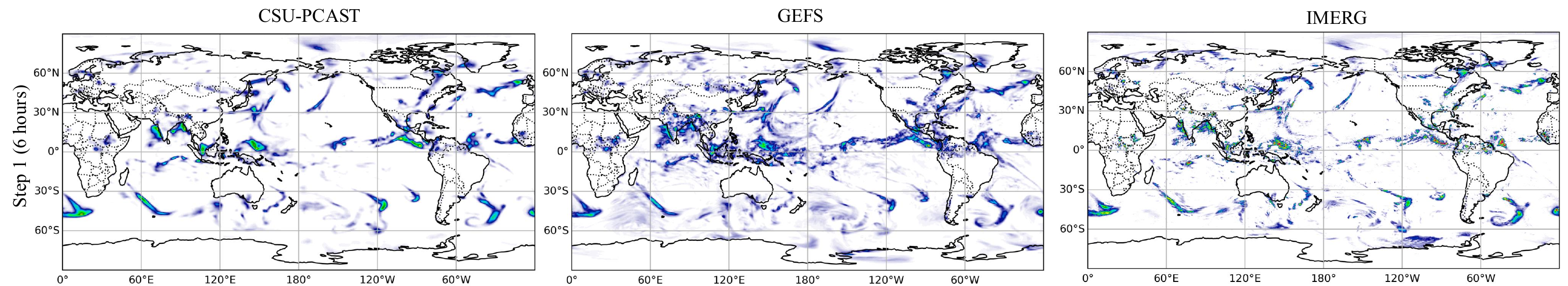}}
\centerline{\includegraphics[width=1\textwidth]{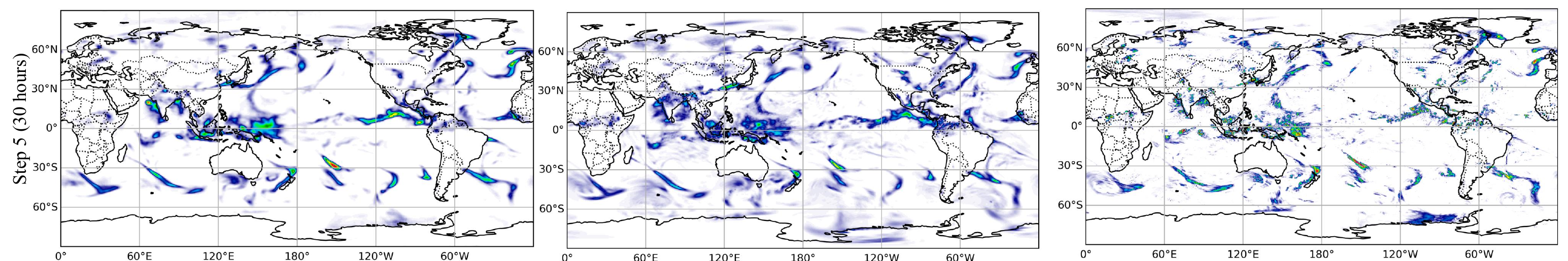}}
\centerline{\includegraphics[width=1\textwidth]{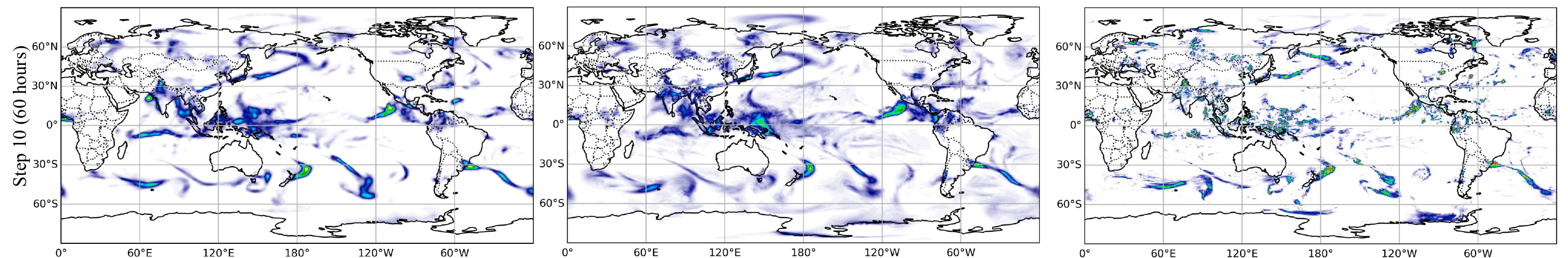}}
\centerline{\includegraphics[width=1\textwidth]{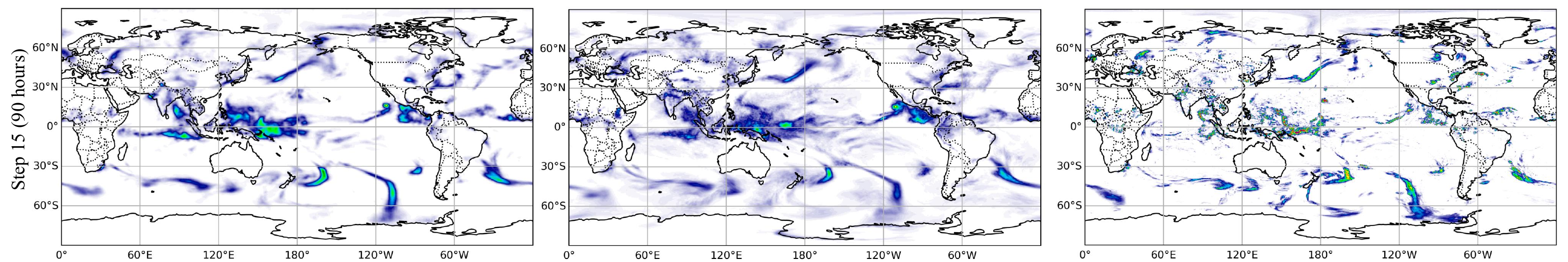}}
\centerline{\includegraphics[width=1\textwidth]{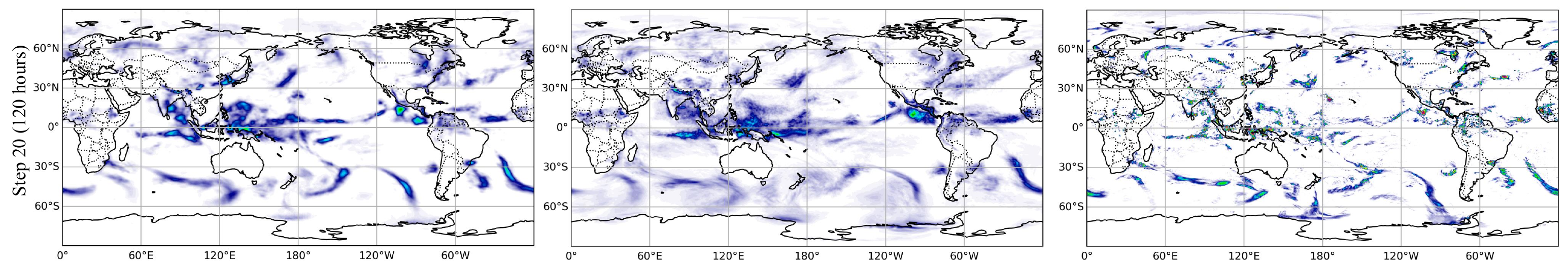}}
\centerline{\includegraphics[width=0.3\textwidth]{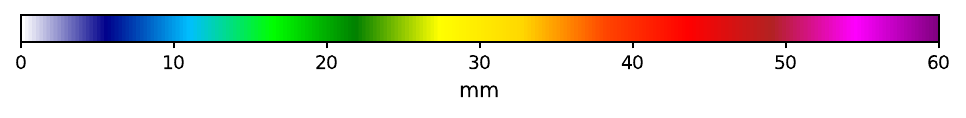}}
\caption{Global 6-h accumulated precipitation forecasts initialized at 00 UTC 6 July 2023. The left and middle columns show 30-member ensemble-mean forecasts from CSU-PCAST and GEFS, respectively, and the right column shows the corresponding IMERG precipitation reference. Rows show forecast steps 1, 5, 10, 15, and 20, corresponding to lead times of 6, 30, 60, 90, and 120 h.}
    \label{fig:precip_comparison}
\end{figure*}

Fig.~\ref{fig:precip_comparison} provides a qualitative example from the CSU-PCAST model trained with IMERG precipitation labels, initialized at 00 UTC 6 July 2023. The CSU-PCAST ensemble mean shows less widespread weak precipitation than GEFS and more closely resembles the intermittent structure of the IMERG reference, consistent with the IMERG-based precipitation supervision used in the main model.

\FloatBarrier

\subsection{Additional CONUS Verification against Stage IV}

\begin{figure*}[htbp]
\centerline{\includegraphics[width=1\textwidth]{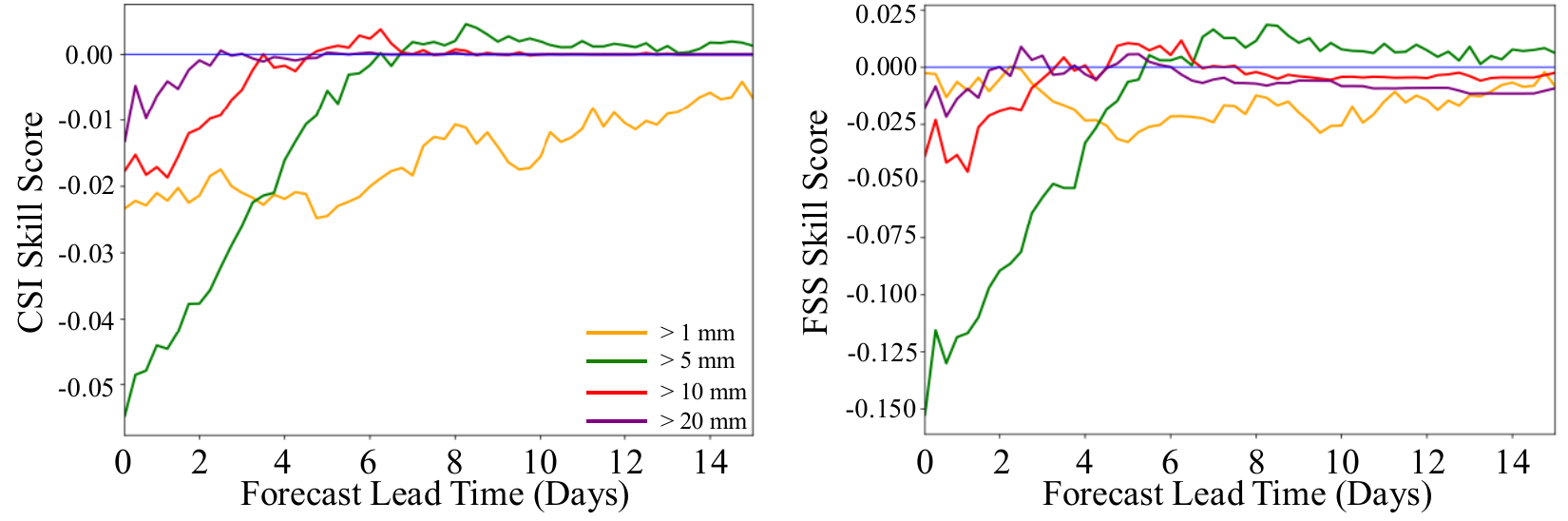}}
\caption{CONUS verification of ensemble-mean precipitation forecasts against Stage IV during the 2023 evaluation period. The left panel shows CSI skill score and the right panel shows FSS skill score for CSU-PCAST relative to GEFS, so positive values indicate better CSU-PCAST performance. Skill scores are evaluated over \(20^\circ\)--\(58^\circ\)N and \(226^\circ\)--\(300^\circ\)E for precipitation thresholds of 1, 5, 10, and 20 mm as a function of forecast lead time.}
\label{fig:conus_skill}
\end{figure*}

\noindent Fig.~\ref{fig:conus_skill} shows an additional CONUS verification of ensemble-mean precipitation forecasts against Stage IV over the 2023 evaluation period. The verification domain covers \(20^\circ\)--\(58^\circ\)N and \(226^\circ\)--\(300^\circ\)E. Skill scores are computed as CSU-PCAST relative to GEFS, so positive values indicate better CSU-PCAST performance. FSS is computed for 6-h precipitation forecasts using a 9-grid-cell neighborhood window at thresholds of 1, 5, 10, and 20 mm. In contrast to the global IMERG-based verification, the CONUS Stage IV results show mixed behavior across thresholds and lead times. For the 1 mm threshold, CSU-PCAST generally has lower CSI and FSS than GEFS throughout most of the forecast range. At the 5 mm threshold, CSU-PCAST performs worse at short lead times but becomes comparable to or slightly better than GEFS after about forecast days 6--7. For the 10 and 20 mm thresholds, the skill differences are generally closer to zero, with brief periods of positive skill but no consistent advantage across the full forecast range. These results indicate that the relative performance of CSU-PCAST is sensitive to the verification region, precipitation reference dataset, threshold, and spatial metric, and that improvements found in the global IMERG-based evaluation do not transfer uniformly to CONUS verification against Stage IV.

\FloatBarrier

\subsection{Latitude-Band Verification}
\label{sec:latband}
\begin{figure*}[htbp]
\centerline{\includegraphics[width=1\textwidth]{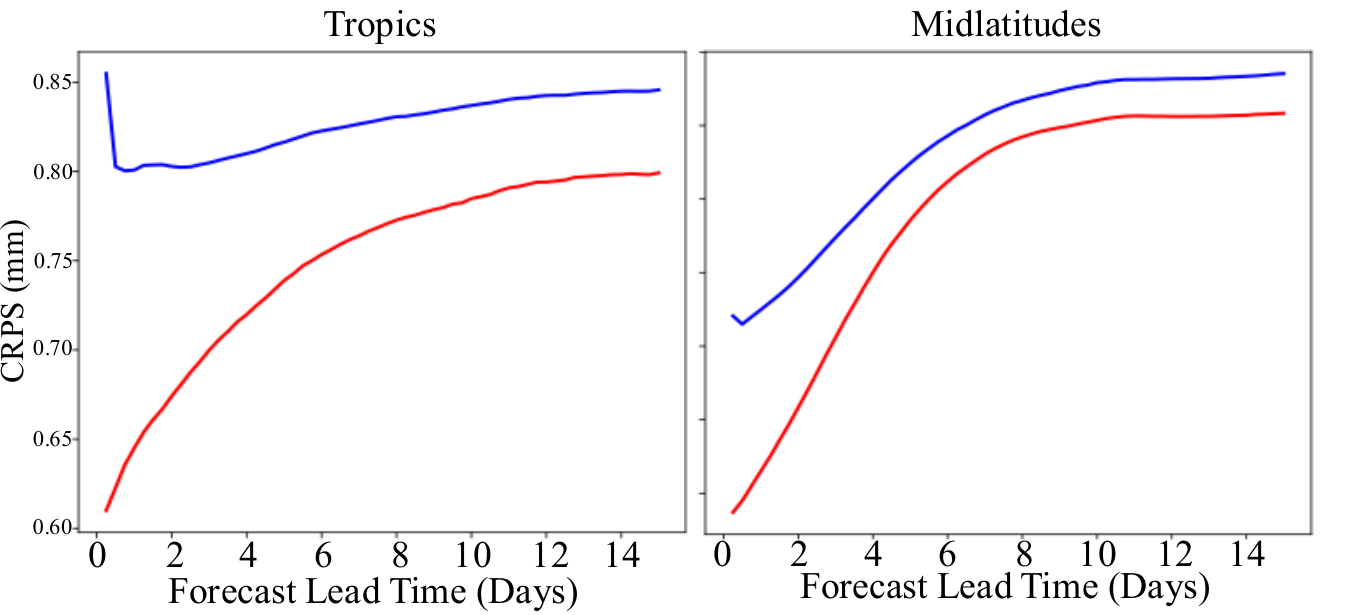}}
\centerline{\includegraphics[width=0.25\textwidth]{IMG/legend.pdf}}
    \caption{CRPS of CSU-PCAST and GEFS precipitation forecasts over the tropics and midlatitudes during the 2023 evaluation period. The tropical region is defined as \(23.5^\circ\)S--\(23.5^\circ\)N, and the midlatitude region includes \(23.5^\circ\)--\(60^\circ\) in both hemispheres. Lower CRPS indicates better probabilistic forecast performance.}
    \label{fig:lat_crps}
\end{figure*}

\noindent To examine whether the relative performance of CSU-PCAST and GEFS depends on precipitation regime, we further separate the global verification into tropical and midlatitude regions. The tropical region is defined as \(23.5^\circ\)S--\(23.5^\circ\)N, and the midlatitude region includes \(23.5^\circ\)--\(60^\circ\) in both hemispheres. Fig.~\ref{fig:lat_crps} shows that CSU-PCAST has lower CRPS than GEFS in both latitude bands throughout the forecast range, indicating improved probabilistic precipitation performance in both tropical and midlatitude regimes.

\begin{figure*}[htbp]
\centerline{\includegraphics[width=1\textwidth]{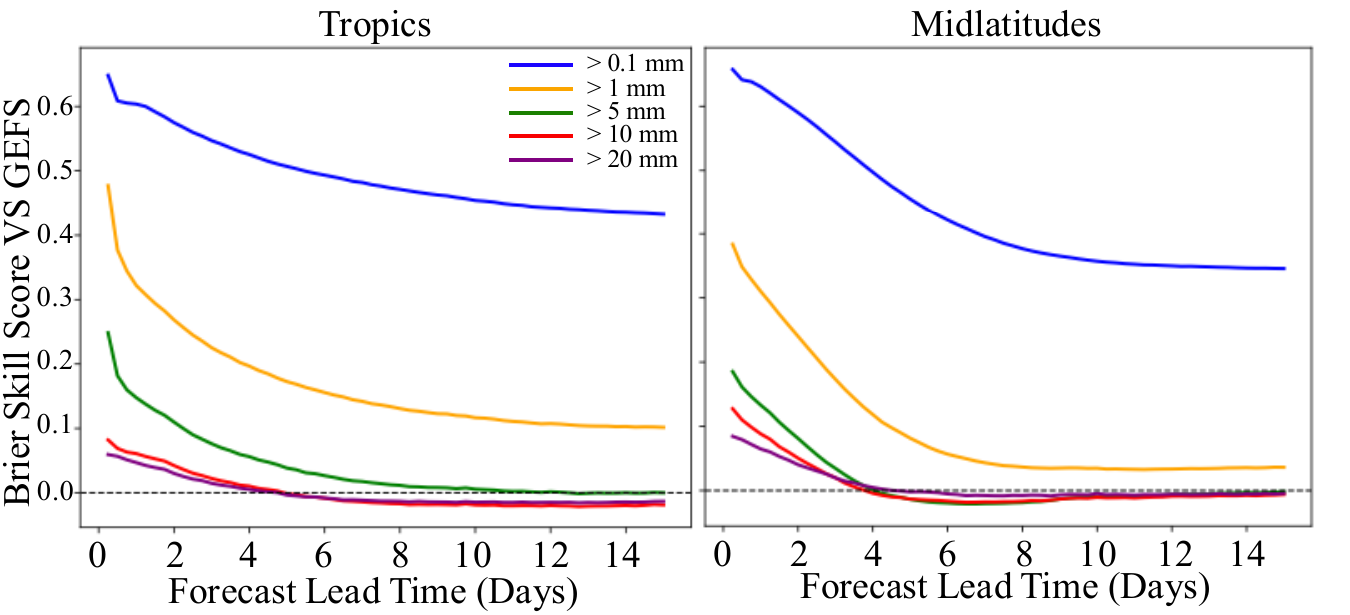}}
    \caption{Brier Skill Score of CSU-PCAST relative to GEFS over the tropics and midlatitudes during the 2023 evaluation period. Positive values indicate improvement over GEFS. Scores are shown for precipitation thresholds of 0.1, 1, 5, 10, and 20 mm as a function of forecast lead time.}
    \label{fig:lat_bs}
\end{figure*}

The Brier Skill Score results in Fig.~\ref{fig:lat_bs} show stronger threshold dependence. CSU-PCAST has positive BSS for light precipitation in both regions, with the largest and most persistent improvements at the 0.1 mm threshold. The skill decreases as the precipitation threshold increases. In the tropics, CSU-PCAST maintains positive BSS for the 1 mm threshold and remains near neutral for the 5 mm threshold at longer leads, while the 10 and 20 mm thresholds become slightly negative after several forecast days. In the midlatitudes, BSS also decreases with threshold and lead time, with heavier thresholds approaching or falling slightly below zero at longer leads.

\begin{figure*}[htbp]
\centerline{\includegraphics[width=1\textwidth]{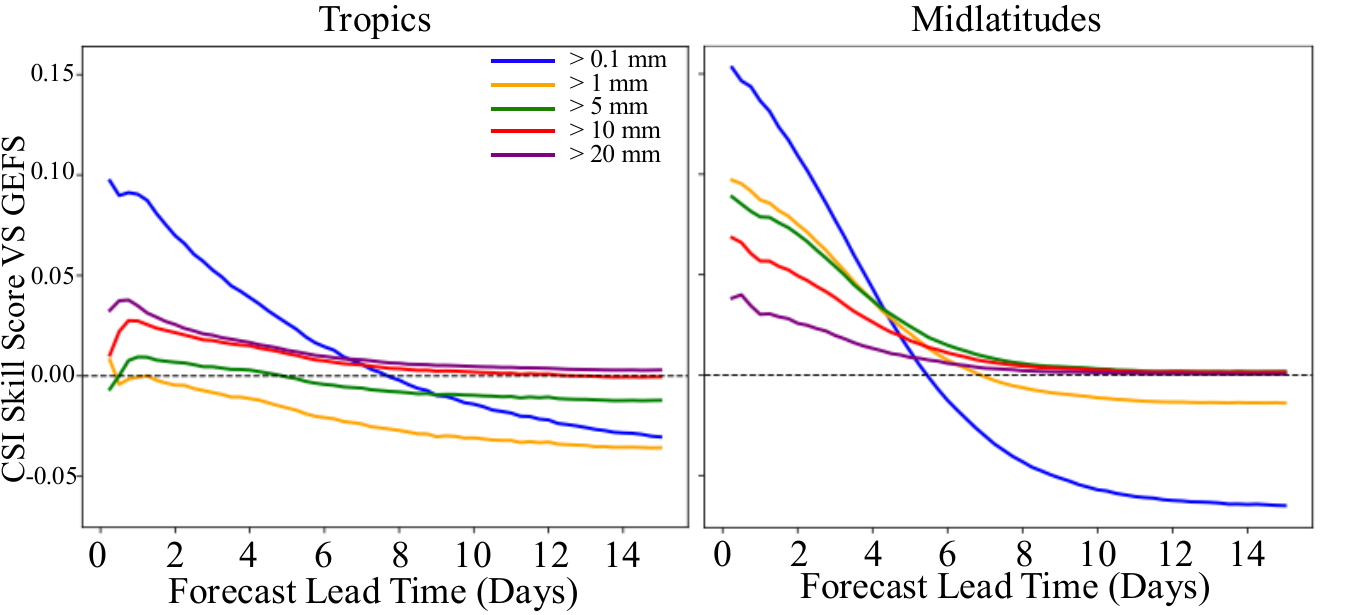}}
    \caption{Member-level CSI skill score of CSU-PCAST relative to GEFS over the tropics and midlatitudes during the 2023 evaluation period. Positive values indicate higher CSI for CSU-PCAST than GEFS. Scores are shown for precipitation thresholds of 0.1, 1, 5, 10, and 20 mm as a function of forecast lead time.}
    \label{fig:lat_csi}
\end{figure*}

Fig.~\ref{fig:lat_csi} shows the member-level CSI skill score of CSU-PCAST relative to GEFS. The CSI skill differs more strongly by latitude band than CRPS. In the tropics, CSU-PCAST shows positive CSI skill for the lightest threshold at short lead times, but this advantage weakens and becomes negative at longer leads; skill at the 1 mm threshold is generally negative, while the 10 and 20 mm thresholds remain closer to neutral or slightly positive. In the midlatitudes, CSU-PCAST shows stronger short-lead CSI improvements across thresholds, but the advantage decreases with lead time, especially for the lightest threshold. These latitude-band diagnostics indicate that the global mean results mask important regional-regime dependence, with CSU-PCAST showing robust CRPS improvements but more mixed categorical skill depending on latitude band, threshold, and lead time.

\FloatBarrier

\subsection{Global Spatial-Activity Diagnostics}
\label{sec:spatial_activity_appendix}

\noindent We further examine the spatial activity of 6-h precipitation forecasts using gradient-based diagnostics. These diagnostics assess whether the forecast fields reproduce the spatial variability and sharp precipitation structures seen in IMERG. The analysis is performed globally over the 2023 evaluation period, with cosine-latitude weighting applied to spatial averages and fractions. All diagnostics are shown as ratios to the corresponding IMERG value, so values closer to one indicate spatial activity more similar to IMERG.

Four metrics are considered. The mean gradient magnitude measures the domain-mean \(|\nabla P|\), reflecting the overall spatial variability of the precipitation field. The 95th-percentile gradient magnitude focuses on the upper tail of \(|\nabla P|\), emphasizing sharper precipitation edges and rainband structures. The gradient per millimeter normalizes the mean gradient magnitude by the mean precipitation amount, reducing the influence of differences in overall precipitation intensity. The high-gradient fraction measures the fraction of grid cells where the forecast gradient exceeds the IMERG 90th-percentile gradient threshold at the same valid time.

\begin{figure*}[htbp]
\centerline{\includegraphics[width=1\textwidth]{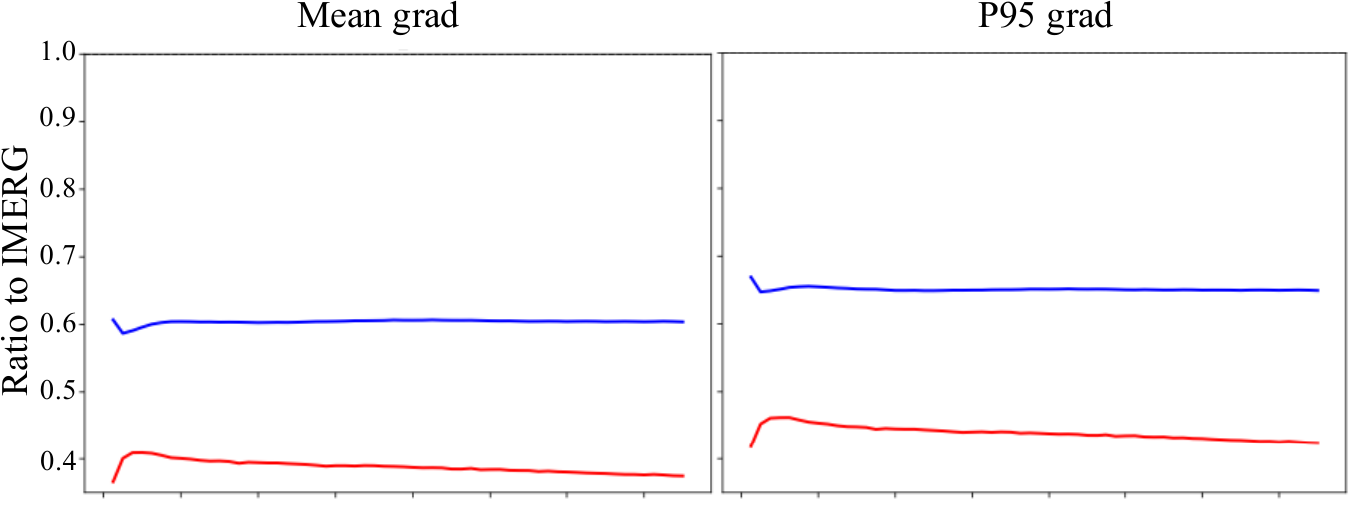}}
\centerline{\includegraphics[width=1\textwidth]{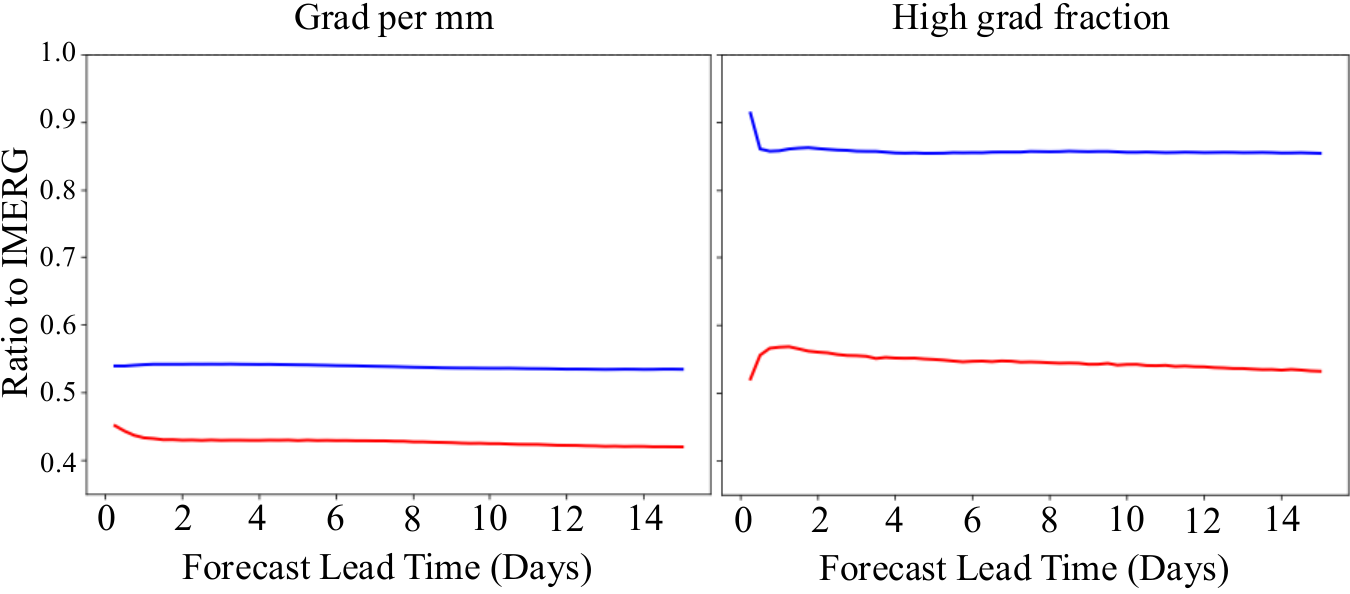}}
\centerline{\includegraphics[width=0.25\textwidth]{IMG/legend.pdf}}
    \caption{Global spatial-activity diagnostics for 6-h precipitation forecasts during the 2023 evaluation period. Values are shown as ratios to IMERG, with values closer to one indicating spatial activity more similar to IMERG. The upper panels show the cosine-latitude-weighted mean precipitation gradient magnitude and the 95th-percentile precipitation gradient magnitude. The lower panels show the gradient magnitude normalized by mean precipitation amount and the high-gradient fraction, defined using the IMERG 90th-percentile gradient magnitude as the threshold.}
    \label{fig:spatial_activity}
\end{figure*}

Fig.~\ref{fig:spatial_activity} shows that both CSU-PCAST and GEFS have weaker spatial activity than IMERG, but the reduction is larger for CSU-PCAST. Averaged across lead times, CSU-PCAST retains about 39\% of the IMERG mean gradient magnitude, compared with about 60\% for GEFS. A similar pattern is found for the 95th-percentile gradient magnitude, where CSU-PCAST retains about 44\% of the IMERG value and GEFS retains about 65\%. The gradient-per-millimeter diagnostic also remains below one for both systems, indicating that the weaker spatial variability is not only a consequence of differences in mean precipitation amount. The high-gradient fraction shows the largest contrast: CSU-PCAST produces high-gradient areas at about 55\% of the IMERG frequency, while GEFS retains about 86\%. These results indicate that CSU-PCAST produces smoother precipitation fields than both IMERG and GEFS, with fewer sharp small-scale precipitation structures.

\FloatBarrier

\subsection{Representative Non-Precipitation Verification}
\label{sec:nonprecip_appendix}

\begin{figure}[htbp]
\centering
\includegraphics[width=0.95\linewidth]{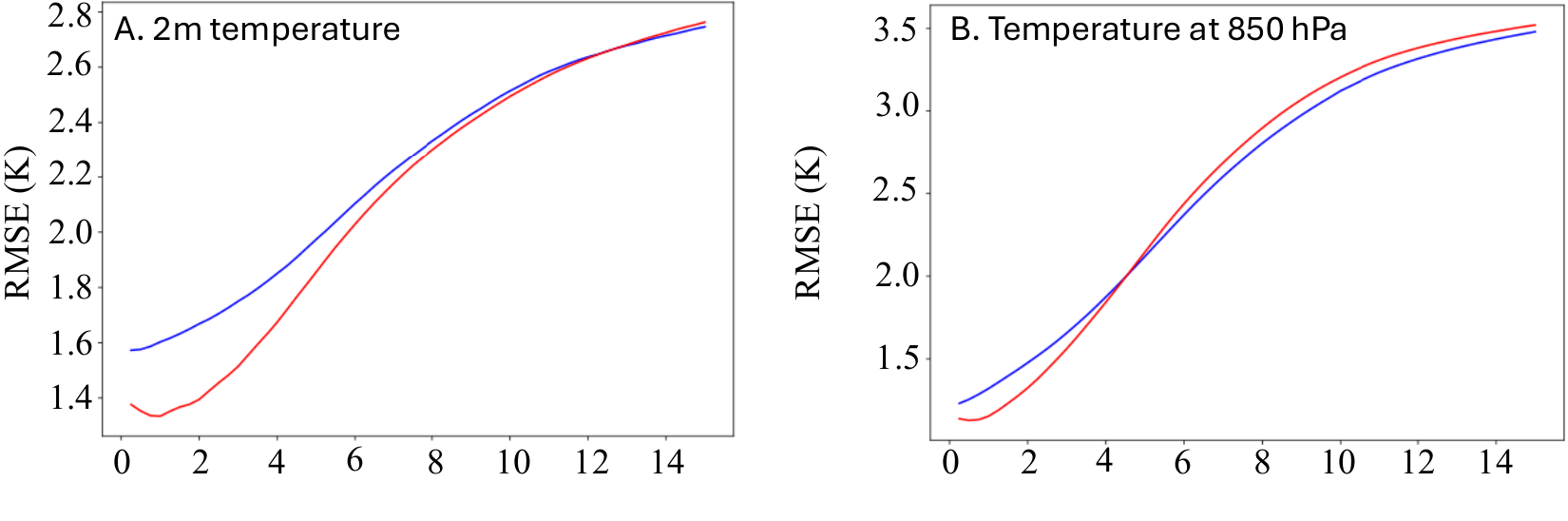}
\includegraphics[width=0.95\linewidth]{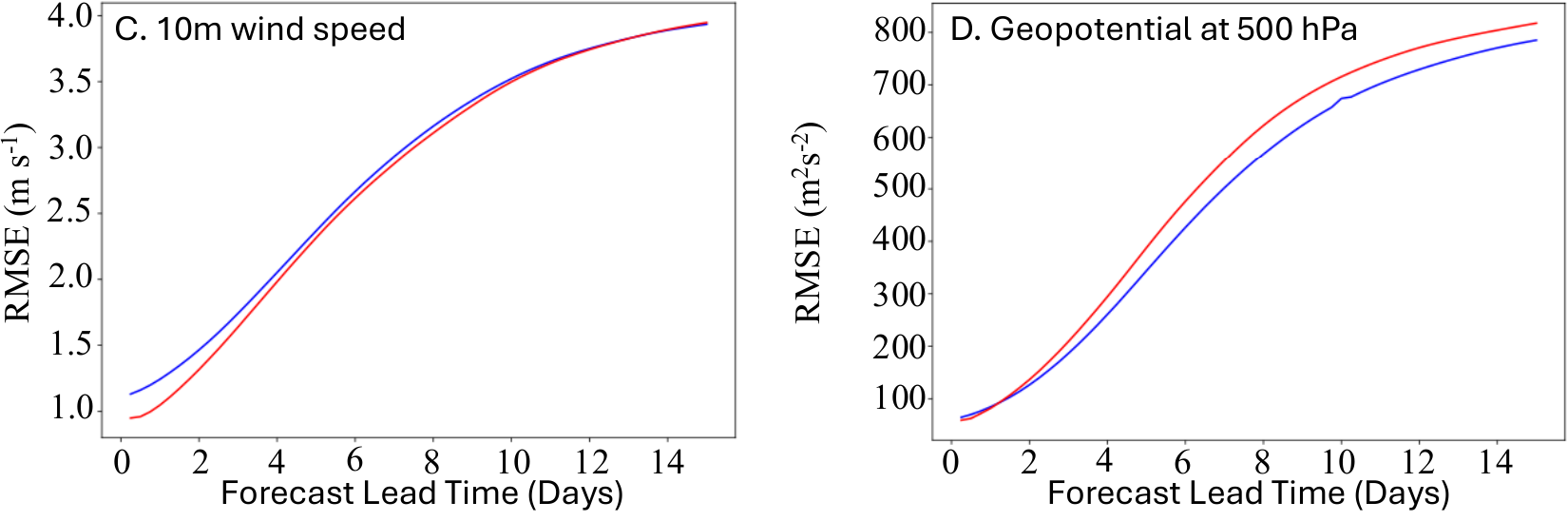}
\includegraphics[width=0.3\linewidth]{IMG/legend.pdf}
\caption{RMSE comparison of the 30-member ensemble-mean forecasts from CSU-PCAST and GEFS for representative non-precipitation variables during the 2023 evaluation period. Panels show 2-m temperature (A), temperature at 850 hPa (B), 10-m wind speed (C), and geopotential at 500 hPa (D). Lower RMSE indicates better forecast performance. CSU-PCAST shows comparable or modestly improved performance for several near-surface variables, while GEFS performs better for 500-hPa geopotential, indicating that non-precipitation skill is variable-dependent.}
\label{fig:nonprecip_rmse}
\end{figure}

\noindent Although this study focuses on precipitation prediction, CSU-PCAST also predicts the non-precipitation atmospheric and surface variables used in the autoregressive rollout. Fig.~\ref{fig:nonprecip_rmse} shows representative RMSE comparisons between CSU-PCAST and GEFS for four non-precipitation variables: 2-m temperature, temperature at 850 hPa, 10-m wind speed, and geopotential at 500 hPa. The results show that CSU-PCAST performs comparably to GEFS for near-surface temperature and wind speed, with lower RMSE at short lead times and similar RMSE at longer leads. For temperature at 850 hPa, CSU-PCAST also shows lower RMSE at the earliest leads, but the advantage decreases with lead time and becomes slightly worse than GEFS at longer leads. For 500-hPa geopotential, CSU-PCAST has larger RMSE than GEFS over much of the forecast range. These results indicate that the non-precipitation branch produces physically useful forecast fields for autoregressive precipitation prediction, but its skill relative to GEFS is variable-dependent and not uniformly better across all atmospheric variables. This is expected to some extent, because the architecture and fine-tuning strategy are designed primarily for precipitation forecasting, and the non-precipitation variables are included mainly to maintain the evolving atmospheric state during autoregressive rollout. In addition, CSU-PCAST uses a limited set of eight pressure levels and precipitation-relevant predictors, rather than the full atmospheric state used by operational NWP systems, which may partly explain the weaker performance for variables such as 500-hPa geopotential.

\FloatBarrier

\end{document}